\documentclass[preprint,12pt]{elsarticle}

\usepackage[ddmmyyyy]{datetime} 
\usepackage{pstricks}

\usepackage{amsmath}
\usepackage{mathtools}
\usepackage{amssymb}
\usepackage{mathrsfs} 
\usepackage{psfrag}
\usepackage[captionskip=12pt]{subfig}
\usepackage{graphicx}
\usepackage{epstopdf}
\usepackage{multirow}
\usepackage{array}
\usepackage{bbm}
\usepackage{enumitem}
\usepackage{xcolor}
\usepackage{amsthm,amssymb}
\usepackage{stmaryrd} 

\usepackage{pstricks}
\newcommand{\DossierRacine}{Images/}

\definecolor{gray80}{gray}{0.8}
\definecolor{gray75}{gray}{0.75}
\definecolor{gray70}{gray}{0.7}
\definecolor{gray60}{gray}{0.6}
\definecolor{gray50}{gray}{0.5}
\definecolor{gray40}{gray}{0.4}

\newcommand{\Eq}[1]{\begin{equation} #1 \end{equation}}
\newcommand{\Tab}[2]{\begin{array}{#1} #2\end{array} }
\newcommand{\Tabu}[2]{\begin{tabular}{#1} #2\end{tabular} }
\newcommand{\BFig}[1]{\begin{figure}\begin{center} #1 \end{center}\end{figure}}
\newcommand{\BPict}[3]{
\begin{pspicture}(#1,#2)
#3
\end{pspicture}
}
\newcommand{\BI}[1]{\begin{itemize}[itemindent=-0.5cm] #1 \end{itemize}}
\newcommand{\BE}[1]{\begin{enumerate} #1 \end{enumerate}}
\newcommand{\PP}[1]{\left( #1 \right)}
\newcommand{\Cr}[1]{\left[ #1 \right]}
\newcommand{\Ac}[1]{\lbrace #1 \rbrace}
\newcommand{\IncGraph}[2]{
\includegraphics[scale=#1]{#2}
}

\newcommand{\cc}{\hspace{1cm}}

\newtheorem{prop}{Proposition}[section]

\numberwithin{equation}{section}

\newcommand{\g}[1]{\boldsymbol{#1}}

\newcommand{\R}[0]{\mathbb{R}}
\newcommand{\N}[0]{\mathbb{N}}
\newcommand{\norme}[1]{\left\Vert  #1 \right\Vert}
\newcommand{\OperateurMinMax}[2]{\underset{#2}{\operatorname{#1}}}

\newcommand{\Mean}[1]{\mathbb{E}\Cr{#1}}

\newcommand{\Normale}[2]{\mathcal{N}\PP{#1,#2}}

\newcommand{\EntreeCodeDeux}[1]{\PP{\varphi^{#1}_1,\g{x}^{#1}_2}}
\newcommand{\ObsEntreeCodeDeux}[0]{\PP{\g{\varphi}_1^{\text{obs}},\g{X}_2^{\text{obs}}}}

\newcommand{\GPCondEcartType}[1]{\sigma^c_{#1}}

\journal{}

\begin{document}

\begin{frontmatter}

\title{Efficient sequential experimental design for surrogate modeling of nested codes}


\author[label1,label2]{Sophie Marque-Pucheu}
\ead{sophie.marque-pucheu@cea.fr }

\author[label1]{Guillaume Perrin}
\author[label3]{Josselin Garnier}

\address[label1]{CEA/DAM/DIF, F-91297, Arpajon, France}
\address[label2]{Laboratoire de Probabilit\'es et Mod\`eles Al\'eatoires,
Universit\'e Paris Diderot, 75205 Paris Cedex 13, France
}
\address[label3]{Centre de Math\'ematiques Appliqu\'ees, Ecole Polytechnique, 91128 Palaiseau Cedex, France}


\begin{abstract}

Thanks to computing power increase, the certification and the conception of complex
systems relies more and more on simulation. To this end, predictive codes are needed, which have generally to be evaluated in a huge number of input points.
When the computational cost of these codes is high, surrogate models are introduced to emulate the response of these codes. In this paper, we consider the situation when the system response can be modeled by two nested computer codes. By two nested computer codes, we mean that some inputs of the second code are outputs of the first code. More precisely, the idea is to propose sequential designs to improve the accuracy of the nested code's predictor by exploiting the nested structure of the codes. In particular, a selection criterion is proposed to allow the modeler to choose the code to call, depending on the expected learning rate and the computational cost of each code. The sequential designs are based on the minimization of the prediction variance, so adaptations of the Gaussian process formalism are proposed for this particular configuration in order to quickly evaluate the mean and the variance of the predictor. The proposed methods are then applied to examples.

\end{abstract}


\end{frontmatter}

\section{Introduction}
\label{PIntroduction}

A lot of industrial issues involve multi-physics phenomena, which can be associated with a series of computer codes. However, when these code networks are used for conception, uncertainty quantification, or risk analysis purposes, they are generally considered as a single code. In that case, all the inputs characterizing the system of interest are gathered in a single input vector, and little attention is paid to the potential intermediate results. When trying to emulate such code networks, this is clearly sub-optimal, as much information is lost in the statistical learning, such that too many evaluations of each code are likely to be required to get a satisfying prediction precision. 

In this paper, we focus on the case of two nested computer codes, which means that the output of the first code is an input of the second code. We assume that these two computer codes are deterministic, but expensive to evaluate. To predict the value of this nested code in a unobserved point, a Bayesian formalism \cite{Robert2007} is adopted in the following. Each computer code is \textit{a priori} modeled by a Gaussian process, and the idea is to identify the posterior distribution of the combination of these two processes given a limited number of evaluations of the two codes. The Gaussian process hypothesis is widely used in computer sciences (\cite{Sacks1989,Santner2003,Rasmussen2006,Kennedy2000,Kennedy2001,Berger2001,Paulo2005,Kleijnen2017}), as it allows a very good trade-off between error control, complexity, and efficiency. The two main issues of this approach, also called Kriging, concern the choice of the statistical properties of the Gaussian processes that are used, and the choice of the points where to evaluate the codes. When a single computer code is considered, several methods exist to add one new point or a batch of new points sequentially to an already existing Design of Experiments (\cite{Sacks1989,Santner2003,Bect2012,Echard2011,Chevalier2014}), in order to minimize the global prediction uncertainty. These methods are generally based on a post-processing of the variance of the code output prediction, which expression can be explicitly derived under mildly restrictive conditions on the mean and the covariance of the prior Gaussian distribution.

The adaptation of these selection criteria to the case of two nested codes is not direct. Indeed, the combination of two Gaussian processes is not Gaussian, such that the prediction uncertainty is much more complicated to estimate. Moreover, if the two codes can be launched separately, the selection criterion has also to indicate which one of the two codes to launch. In that prospect, the first objective of this paper is to propose several adaptations of the Gaussian Process formalism to the nested case, in order to be able to evaluate the two first statistical moments of the code output predictor quickly. Then, original sequential selection criteria are introduced, which try to exploit as much as possible the nested structure of the studied codes. In particular, these criteria are able to integrate the fact that the computational cost associated with the evaluation of each code can be different.

The outline of this paper is the following. Section \ref{sec2} presents the theoretical framework of the Gaussian process-based surrogate models, its generalization to the nested case, and introduces several selection criteria based on the prediction variance to reduce the prediction uncertainty sequentially. Section \ref{sec3} introduces a series of simplifications to allow a quick evaluation of the prediction variance. In section \ref{sec4}, the presented methods are eventually applied to two examples.\\

The proofs of the results that will be presented in the following sections have been moved to the appendix.

\section{Surrogate modeling for two nested computer codes}
\label{sec2}

\subsection{Notations}
\label{sec21}

In this paper, the following notations will be adopted:

\begin{itemize}
\item $x,y$ correspond to scalars.
\item $\g{x},\g{y}$ correspond to vectors.
\item $\g{X},\g{Y}$ correspond to matrices.
\item The entries of a vector $\g{x}$ are denoted by $(\g{x})_i$, whereas the entries of a matrix $\g{X}$ are denoted by $(\g{X})_{ij}$.
\item $\g{X}^T$ denotes the transpose of a matrix $\g{X}$.
\item $\mathcal{N}(\g{x},\g{X})$ corresponds to the multidimensional Gaussian distribution, whose mean vector and covariance matrix are respectively given by $\g{x}$ and $\g{X}$.
\item $\text{GP}(m,k)$ corresponds to the distribution of a Gaussian process whose mean function is $m$, and whose covariance function is $k$.
\item $\mathbb{E}\Cr{\cdot}$ and $\mathbb{V}(\cdot)$ are the mathematical expectation and the variance respectively.
\item For all real-valued functions $y$ and $z$ that are square integrable on $\mathbb{X}$, $(\cdot,\cdot)_{\mathbb{X}}$ and $\norme{\cdot}_{\mathbb{X}}$ denote respectively the classical scalar product and norm in the space of square integrable real-valued functions on $\mathbb{X}$:

\Eq{(y,z)_{\mathbb{X}}:=\int_{\mathbb{X}}y(\g{x})z(\g{x})d\g{x}, \ \ \norme{y}^2_{\mathbb{X}}:=(y,y)_{\mathbb{X}}.} 

\end{itemize}

\subsection{General framework}
\label{sec22}

Let $\mathcal{S}$ be a system that is characterized by a vector of input parameters, $\g{x}_{\text{nest}}\in\mathbb{X}_{\text{nest}}$. Let $y_{\text{nest}}: \mathbb{X}_{\text{nest}} \rightarrow \R$ be a deterministic mapping that is used to analyze the studied system. In this paper, we focus on the case where the function $\g{x}_{\text{nest}}\mapsto y_{\text{nest}}(\g{x}_{\text{nest}})$ can be modeled by two nested codes. Two quantities of interest, $y_1$ and $y_2$, are thus introduced to characterize these two codes, which are supposed to be two real-valued continuous functions on their respective definition domains $\mathbb{X}_1$ and $\R\times \mathbb{X}_2$. Given these two functions, the nested code is defined as follows:

\Eq{
\Tab{c}{
 \\
\\
\g{x}_1\in \mathbb{X}_1
}
\
\Tab{c}{
 \\
\\
\rightarrow
}
\
\Tab{c}{
\g{x}_2\in \mathbb{X}_2 \\
\\
y_1(\g{x}_1) \in\R
}
\
\Tab{c}{
 \searrow \\
\nearrow
}
\
\Tab{c}{
 y_{\text{nest}}(\g{x}_{\text{nest}}):=y_2(y_1(\g{x}_1),\g{x}_2) \in\R,
}
\label{eqPhenomeneCouple}
}

\noindent where $\g{x}_{\text{nest}}:=(\g{x}_1,\g{x}_2)\in\mathbb{X}_{\text{nest}}=\mathbb{X}_1\times \mathbb{X}_2$. The sets $\mathbb{X}_1$ and $\mathbb{X}_2$ are moreover supposed to be two compact subsets of $\R^{d_1}$ and $\R^{d_2}$ respectively, where $d_1$ and $d_2$ are two positive integers. In theory, the definition domains may be unbounded, but the reduction to compact sets enables the square integrability of $y_{\text{nest}}$ on $\mathbb{X}_{\text{nest}}$.

Given a limited number of evaluations of the functions $\g{x}_1\mapsto y_1(\g{x}_1)$ and $\EntreeCodeDeux{}\mapsto y_2\EntreeCodeDeux{}$, the objective is to build a stochastic predictor of $y_{\text{nest}}$ with the following properties:
\BI{
\item its mean is as close as possible to the real output of the nested code, that is, the bias is small,
\item its uncertainty (given by its variance) is as small as possible.
}
In other words, the mean square error of the stochastic predictor has to be small.

\subsection{Gaussian process-based surrogate models}
\label{sec23}
The Gaussian process regression (GPR), or Kriging, is a technique that is widely used to replace an expensive computer code by a surrogate model, that is to say a fast to evaluate mathematical function. The GPR is based on the assumption that the two code outputs, $y_1$ and $y_2$, can be seen as the sample paths of two stochastic processes, $\widehat{y}_1$ and $\widehat{y}_2$, which are supposed to be Gaussian for the sake of tractability:

\Eq{\widehat{y}_i\sim \text{GP}(\mu_i,C_i), \ \ i\in\Ac{1,2},\label{priorGP}}

\noindent where for all $1\leq i\leq 2$, $\mu_i$ and $C_i$ denote respectively the mean and the covariance functions of $\widehat{y}_i$. \\
Let $\g{x}_1^{(1)},\ldots,\g{x}_1^{(N_1)}$ be $N_1$
elements of $\mathbb{X}_1$ and $\PP{\varphi_1^{(1)},\g{x}_2^{(1)}},\ldots,(\varphi_1^{(N_2)},\g{x}_2^{(N_2)})$ be $N_2$ elements of $\R\times \mathbb{X}_2$. Denoting by

\Eq{\g{y}^{\text{obs}}_1:=(y_1(\g{x}_1^{(1)}),\ldots,y_1(\g{x}_1^{(N_1)})), \ \ \ \g{y}^{\text{obs}}_2:=(y_2(\varphi_1^{(1)},\g{x}_2^{(1)}),\ldots,y_2(\varphi_1^{(N_2)},\g{x}_2^{(N_2)})),}

\noindent the vectors that gather the evaluations of $y_1$ and $y_2$ in these points, it can be shown that:

\Eq{\widehat{y}^c_i:=\widehat{y}_i \ \vert \ {\g{y}}^{\text{obs}}_i \ \sim \ \text{GP}(\mu_i^{c},C_i^{c}),\label{predictGP}}

\noindent and we refer to \cite{Sacks1989,Santner2003} for further details about the expressions of conditioned mean functions, $\mu_i^{c}$, and conditioned covariance functions, $C_i^{c}$.

According to Eq. \eqref{eqPhenomeneCouple}, the nested code, $\g{x}_{\text{nest}}\mapsto y_{\text{nest}}(\g{x}_{\text{nest}})$, can thus be seen as a particular realization of the conditioned process $\widehat{y}^c_{\text{nest}}$, such that for all $(\g{x}_1,\g{x}_2)\in\mathbb{X}_1\times \mathbb{X}_2$,

\Eq{\widehat{y}_{\text{nest}}^c(\g{x}_1,\g{x}_2):=\widehat{y}^c_2(\widehat{y}^c_1(\g{x}_1),\g{x}_2).} 

Under this Gaussian formalism, the best prediction of $y_{\text{nest}}$ in any unobserved point $\g{x}_{\text{nest}}=(\g{x}_1,\g{x}_2)$ in $\mathbb{X}_1\times \mathbb{X}_2$ is given by the mean value of $\widehat{y}_{\text{nest}}^c(\g{x}_1,\g{x}_2)$, whereas its variance can be used to characterize the trust we can put in that prediction. As explained in Introduction, there is no reason for $\widehat{y}_{\text{nest}}^c$ to be Gaussian, but according to Proposition \ref{prop1}, the first- and second-order moments can be obtained by computing two one-dimensional integrals with respect to a Gaussian measure. This can be done by quadrature rules or by Monte-Carlo methods (\cite{Baker1977}).

\begin{prop}
\label{prop1}
For all $(\g{x}_1,\g{x}_2)\in\mathbb{X}_1\times \mathbb{X}_2$, if $\xi\sim\mathcal{N}(0,1)$, then:

\Eq{\mathbb{E}\Cr{\widehat{y}_{\text{nest}}^c(\g{x}_1,\g{x}_2)}=\mathbb{E}\Cr{\mu_2^c(\mu_1^c(\g{x}_1)+\sigma_1^c(\g{x}_1)\xi,\g{x}_2)},\label{moment1Nested}}

\Eq{\mathbb{E}\Cr{\PP{\widehat{y}_{\text{nest}}^c(\g{x}_1,\g{x}_2)}^2}=\mathbb{E}\Cr{
\begin{split}
& \Ac{\mu_2^c(\mu_1^c(\g{x}_1)+\sigma_1^c(\g{x}_1)\xi,\g{x}_2)}^2 \\ & +\Ac{\sigma_2^c(\mu_1^c(\g{x}_1)+\sigma_1^c(\g{x}_1)\xi,\g{x}_2)}^2
\end{split}},\label{moment2Nested}}

\noindent where for all i in $\Ac{1,2}$, $(\sigma_i^c\PP{\g{x}_i})^2=C_i^c\PP{\g{x}_i,\g{x}_i}$. 
\end{prop}

\subsection{Parametric representations of the mean and covariance functions}
\label{sec24}

As explained in Introduction, the relevance of the Gaussian process predictor strongly depends on the definitions of $\mu_i$ and $C_i$. When the maximal information about $y_i$ is a finite set of evaluations, these functions are generally chosen in general parametric families. In this paper, functions $C_i$ are supposed to be two elements of the Mat\'ern-5/2 class (see \cite{Santner2003,Stein1999} for further details about classical parametric expressions for $C_i$), with $\g{\theta}_i$ be the hyper-parameters that characterize these covariance functions, whereas linear representations are considered for the mean functions,

\Eq{\mu_i=\g{h}_i^T\g{\beta}_i,}

\noindent where $\g{h}_i$ is a given $M_i$-dimensional vector of functions (see \cite{PerrinJCP2017} for further details on the choice of the basis functions). In the following, the framework of the "Universal Kriging" is adopted, which consists in:

\begin{itemize}
\item assuming an (improper) uniform distribution for $\g{\beta}_i$,
\item conditioning all the results by the maximum likelihood estimate of $\g{\theta}_i$,
\item integrating over $\g{\beta}_i$ the conditioned distribution of $\widehat{y}_i$.
\end{itemize}

In that case, the distribution of $\widehat{y}^c_i$, which is defined by Eq. \ref{predictGP} is Gaussian, and its statistical moments can explicitly be derived (see \cite{Sacks1989,Bichon2008,Bect2012,PerrinJCP2017}).

\subsection{Sequential designs for the improvement of Gaussian process predictors}
\label{sec25}

The relevance of the predictor $\widehat{y}_{\text{nest}}^c$ strongly depends on the space filling properties of the sets gathering the inputs of the available observations of $y_1$ and $y_2$, which are generally called Designs of Experiments (DoE). Space-filling Latin Hypercube Samplings (LHS) or quasi-Monte-Carlo samplings are generally chosen to define such \textit{a priori} DoE (\cite{Fang2003,Fang2006,PerrinSFDS2017}). The relevance of the predictor can then be improved by adding new points to an already existing DoE, as the higher the values of $N_1$ and $N_2$, the more chance there is for $\norme{\mathbb{E}\Cr{\widehat{y}_{\text{nest}}^c}-{y}_{\text{nest}}}_{\mathbb{X}_{\text{nest}}}^2$ to be small. 

\cc

In the case of a single code, most of the existing selection criteria to add a new point are based on the minimization of a quantity associated with the predictor variance, such as its integral over the input domain for instance \cite{Sacks1989,Santner2003,Echard2011,Bect2012,Chevalier2014,Perrin2016,Hu2017,Gramacy2012Tech}. Indeed, if $\widehat{z}$ is a Gaussian process that is indexed by $\g{x}$ in $\mathbb{X}$, and if we denote by $k$ its covariance function, the variance of the conditioned random variable $\widehat{z}(\g{x}) \ \vert \ \widehat{z}(\g{x}^{\text{new}})$, where $\g{x}$ and $\g{x}^{\text{new}}$ are any elements of $\mathbb{X}$, is given by:

\Eq{k(\g{x},\g{x})-k(\g{x},\g{x}^{\text{new}})^2/k(\g{x}^{\text{new}},\g{x}^{\text{new}}),}

\noindent such that it does not depend on the (unknown) value of $\widehat{z}(\g{x}^{\text{new}})$. To minimize the global uncertainty over $\widehat{z}$ at a reduced computational cost, a natural approach would consist in searching the value of $\g{x}^{\text{new}}$ such that 

\Eq{\int_{\mathbb{X}}\Ac{k(\g{x},\g{x})-k(\g{x},\g{x}^{\text{new}})^2/k(\g{x}^{\text{new}},\g{x}^{\text{new}})}d\g{x}}

\noindent is minimal (under the condition that this integral exists).

\cc

In the nested case, we also have to choose on which code to add a new observation point. To this end, let $\tau_1$ and $\tau_2$ be the numerical costs (in CPU time for instance) that are associated with the evaluations of $y_1$ and $y_2$ respectively. For the sake of simplicity, we assume that these numerical costs are independent on the value of the input parameters, and that they are \textit{a priori} known. Two selection criteria are eventually proposed to optimize the relevance of the Gaussian process predictor sequentially. To simplify the reading, the following notation is proposed:

\Eq{(\widetilde{\g{x}}_i,\widetilde{\mathbb{X}}_i):=
\left\lbrace 
\begin{split}
& (\g{x}_1,\mathbb{X}_1) \ \text{if} \ i=1, \\
& (\EntreeCodeDeux{},\R\times \mathbb{X}_2) \ \text{if} \ i=2, \\
& ((\g{x}_1,\g{x}_2),\mathbb{X}_1\times \mathbb{X}_2) \ \text{if} \ i=3,
\end{split}
\right. 
}

\noindent and we denote by $\mathbb{V}(\widehat{y}^c_{\text{nest}}(\g{x}_{\text{nest}}) \vert \widetilde{\g{x}}_i)$ the variance of $\widehat{y}^c_{\text{nest}}(\g{x}_{\text{nest}})$ under the hypothesis that the code(s) corresponding to the new point $\widetilde{\g{x}}_i$ is(are) evaluated in this point (in practice, we remind that these code evaluations are not required for the estimation of this variance).

\BI{

\item First, the chained I-optimal criterion selects the best point in $\mathbb{X}_1 \times \mathbb{X}_2 $ to minimize the integrated variance of the predictor of the nested code: 
\Eq{
\widetilde{\g{x}}_3^{\text{new}}=\OperateurMinMax{arg min}{\widetilde{\g{x}}_3\in \widetilde{\mathbb{X}}_3} \int_{\mathbb{X}_{\text{nest}}}\mathbb{V}(\widehat{y}^c_{\text{nest}}(\g{x}_{\text{nest}}) \vert \widetilde{\g{x}}_3)d\g{x}_{\text{nest}}.\label{selectopt2}
}

Such a criterion is \textit{a priori} adapted to the case when it is not possible to run independently the codes 1 and 2.

\item Secondly, the best I-optimal criterion selects the best among the candidates in $\mathbb{X}_1$ and $\mathbb{X}_2 $ in order to maximize the decrease per unit of computational cost of the integrated predictor variance of the nested code: 
\Eq{
(i^{\text{new}},\widetilde{\g{x}}_{i^{\text{new}}}^{\text{new}})=\OperateurMinMax{arg max}{\widetilde{\g{x}}_i\in \widetilde{\mathbb{X}}_i,\ i\in\Ac{1,2}}
\dfrac{1}{\tau_i}
\times \int_{\mathbb{X}_{\text{nest}}}
\Cr{
\mathbb{V}\PP{\widehat{y}^c_{\text{nest}}(\g{x}_{\text{nest}}) }
-\mathbb{V}\PP{\widehat{y}^c_{\text{nest}}(\g{x}_{\text{nest}}) \vert \widetilde{\g{x}}_i}
}
d\g{x}_{\text{nest}}.\label{selectopt1}
}

\noindent In that case, the difference in the computational costs is taken into account, and a linear expected improvement per unit of computational cost is assumed for the sake of simplicity.

}

\section{Fast evaluation of the prediction variance}
\label{sec3}

As explained in Section \ref{sec25}, to choose the position of the new point, for each potential value of $\widetilde{\g{x}}_i$ in $\widetilde{\mathbb{X}}_i$, we need to compute the value of $\text{Var}(\widehat{y}^c_{\text{nest}}(\g{x}_{\text{nest}}) \vert \widetilde{\g{x}}_i)$ for all $\g{x}_{\text{nest}}$ in $\mathbb{X}_{\text{nest}}$. If quadrature rules or Monte Carlo approaches are used to evaluate this variance, as it is proposed in Section \ref{sec23}, the optimization procedure quickly becomes extremely demanding, even if discretized approximations of the optimization problem defined by Eqs. \eqref{selectopt1} and \eqref{selectopt2} are considered, that is to say where the integral over $\mathbb{X}_{\text{nest}}$ is replaced by an empirical mean over any $N_{\text{nest}}$-dimensional set of randomly chosen points of $\mathbb{X}_{\text{nest}}$. To circumvent this problem, we present in this section several approaches to make the evaluation of $\text{Var}(\widehat{y}^c_{\text{nest}}(\g{x}_{\text{nest}}) \vert \widetilde{\g{x}}_i)$ explicit, and therefore extremely fast to evaluate.

\subsection{Explicit derivation of the two first statistical moments of the nested code predictor}
\label{sec31}

\begin{prop}
\label{prop2}

Using the notations of the Universal Kriging framework that is introduced in Section \ref{sec24}, and denoting by $g$ the family of functions such that $g\PP{x,\g{\alpha}} := x^{(\g{\alpha})_1 } \exp\Cr{(\g{\alpha})_2 x+(\g{\alpha})_3 x^2},\ \g{\alpha } \in \N \times \R^2$ if:

\BE{
\item for $1\leq k\leq M_2$ the mean function $\PP{\g{h}_2}_k$ is of the form:
\Eq{(\g{h}_2(\EntreeCodeDeux{})_k=m_k(\g{x}_2)\ g\PP{\varphi_1,\g{\alpha}_k},}
where $m_k$ is a deterministic function from $\mathbb{X}_2$ to $\R$ and $\g{\alpha}_k \in \N \times \R^2$ is such that $2(\g{\alpha}_k)_3 C_1^c(\g{x}_1,\g{x}_1)<1$ for all $\g{x}_1 \in \mathbb{X}_1$, 
\item the covariance function $C_2$ is an element of the Gaussian class or corresponds to the covariance function of any derivative of a zero-mean process with covariance function of the Gaussian class,
}
then the conditional moments of order 1 and 2 of $\widehat{y}_{\text{nest}}^c(\g{x}_1,\g{x}_2)$, which are defined by Eqs. (\ref{moment1Nested}) and (\ref{moment2Nested}) can be calculated analytically.

\end{prop}

In other words, if the prior of the Gaussian process modeling the function $y_2$ can be seen as any derivative of a Gaussian process with a trend which is a linear combination of products of polynomials by exponentials of order less than 2, and a covariance function of the Gaussian class, then conditionally to some integration criteria, the moments of order 1 and 2 of the coupling of the predictors of the two codes can be computed explicitly at a reduced cost. However, the approach cannot be generalized to the coupling of more than two codes.

\subsection{Linearized approach}
\label{sec32}

In the cases where the conditions for Proposition \ref{prop2} are not fulfilled (or if more than two codes were considered), another approach is proposed in this section, which is based on a linearization of the process modeling the nested code. Indeed, for $i\in\Ac{1,2}$, let $\varepsilon_i^c$ be the Gaussian process such that:

\Eq{\widehat{y}_i^c=\mu^c_i+\varepsilon_i^c.}

By construction, $\varepsilon_i^c$ is the residual prediction uncertainty once $\widehat{y}_i$ has been conditioned by $N_i$ evaluations of $y_i$. We remind that these two Gaussian processes are statistically independent. Under the condition that $N_1$ is not too small compared to the complexity of $y_1$, it is therefore reasonable to assume that $\varepsilon^c_1$ is small compared to $\mu_1^c$. 

\begin{prop}
If:
\BE{
\item the predictor of two nested computer codes can be written $\widehat{y}_{\text{nest}}^c(\g{x}_1,\g{x}_2):=\widehat{y}^c_2(\widehat{y}^c_1(\g{x}_1),\g{x}_2)$, 
where $\widehat{y}_i^c$ are Gaussian processes which can be written as 
$\widehat{y}_i^c=\mu^c_i+\varepsilon_i^c$ where $\varepsilon_i^c \sim \text{GP}\PP{0,C_i^c},\quad i\in\Ac{1,2}$,
\item and $\varepsilon_1^c$ is small enough for the linearization to be valid,
}
then the predictor of the two nested computer codes can be defined as a Gaussian process with the following mean and covariance functions:
\Eq{
\Tab{c}{
\mu_{\text{nest}}^c=\mu_{2}^c(\mu_{1}^c(\g{x}_1),\g{x}_2)
\\[10pt]
\begin{split}
C^c_{\text{nest}}((\g{x}_1,\g{x}_2),(\g{x}_1',\g{x}_2')) & =C_2^c((\mu_1^c(\g{x}_1),\g{x}_2),(\mu_1^c(\g{x}_1'),\g{x}_2')) \\ + & \dfrac{\partial \mu_2^c}{\partial \varphi_1}(\mu_1^c(\g{x}_1),\g{x}_2)\dfrac{\partial \mu_2^c}{\partial \varphi_1}(\mu_1^c(\g{x}_1'),\g{x}_2') C_1^c(\g{x}_1,\g{x}_1').
\end{split}
}
\label{prop3eq1}
}
\label{prop3}
\end{prop}

Hence, thanks to the proposed linearization, the variance of $\widehat{y}_{\text{nest}}^c(\g{x}_{\text{nest}})$ but also the one of $\widehat{y}^c_{\text{nest}}(\g{x}_{\text{nest}}) \vert \widetilde{\g{x}}_i$ can explicitly be derived for all $(\g{x}_{\text{nest}},\widetilde{\g{x}}_i)$ in $\mathbb{X}_{\text{nest}}\times \widetilde{\mathbb{X}}_i$. Under the condition that the linearization is valid, this approach can be applied to configurations with more than two nested codes.

However it can be inferred from equation \eqref{prop3eq1} that the variance depends on $\g{y}_1^{\text{obs}}$ through $\mu_1^c$. To circumvent this problem for the evaluation of the forward variance in the sequential designs, we assume that a candidate $\g{x}_1$ is associated with the current estimate of the output of the first code $\mu_1^c\PP{\g{x}_1}$, in accordance with the Kriging Believer strategy proposed in \cite{Ginsbourger2010}.

\section{Applications}
\label{sec4}

The previously proposed methods are applied to two examples: an analytical one-dimensional one and a multidimensional one.

\subsection{Characteristics of the examples}

\subsubsection{Analytical example}

In the analytical example the properties of the Gaussian process mean functions and of the codes are: 
\Eq{
\g{h}_1\PP{x_1}=\left[
\Tab{c}{
1\\[3pt]
x_1\\[3pt]
x_1^2
}
\right],
\quad 
\g{\beta}_1^*=\left[
\Tab{c}{
-2\\
0.25\\
0.0625
}
\right], 
\quad 
y_1\PP{x_1}=\g{h}_1\PP{x_1}^T\g{\beta}_1^*-0.25\cos\PP{2\pi x_1},
}

\Eq{
\g{h}_2\PP{\varphi_1}=\left[
\Tab{c}{
1\\[3pt]
\varphi_1\\[3pt]
\varphi_1^2\\[3pt]
\varphi_1^3
}
\right],
\qquad 
\g{\beta}_2^*=\left[
\Tab{c}{
6\\
-5\\
-2\\
1
}
\right], 
\qquad 
y_2\PP{\varphi_1}=\g{h}_2\PP{\varphi_1}^T\g{\beta}_2^* -0.25\cos\PP{2\pi \varphi_1},
}

\noindent where $x_1 \in \Cr{-7,7}$. In this example $\mathbb{X}_2=\emptyset$.

\hspace{0cm}

Figure \ref{fig0} shows the variations of the outputs of the codes 1, 2 and nested. The codes 1 and 2 outputs are relatively smooth compared with the one of the nested code. The amplitude of the variations is strongly non-stationary for the nested code.

\newcommand{\taillefigure}{0.4}

\BFig{
\vspace{-0.7cm}
\Tabu{cc}{
\vspace{-0.7cm}
\psfrag{Titre}[c][c][1]{}
\psfrag{xlab}[c][c][1]{$\g{x}_1$}
\psfrag{ylab}[c][c][1]{$y_1$}
\subfloat[Code 1]{
\IncGraph{\taillefigure}{\DossierRacine 25_ExemplePoly_TraceFonctions/1_Figures/0_TraceFonctions-1.eps}}
&
\psfrag{Titre}[c][c][1]{}
\psfrag{xlab}[c][c][1]{$\g{x}_2$}
\psfrag{ylab}[c][c][1]{$y_2$}
\subfloat[Code 2]{
\IncGraph{\taillefigure}{\DossierRacine 25_ExemplePoly_TraceFonctions/1_Figures/0_TraceFonctions-2.eps}}
\\

\multicolumn{2}{c}{
\psfrag{Titre}[c][c][1]{}
\psfrag{xlab}[c][c][1]{$\g{x}_{\text{nest}}$}
\psfrag{ylab}[c][c][1]{$y_{\text{nest}}$}
\subfloat[Nested code]{
\IncGraph{\taillefigure}{\DossierRacine 25_ExemplePoly_TraceFonctions/1_Figures/0_TraceFonctions-3.eps}}
}
}
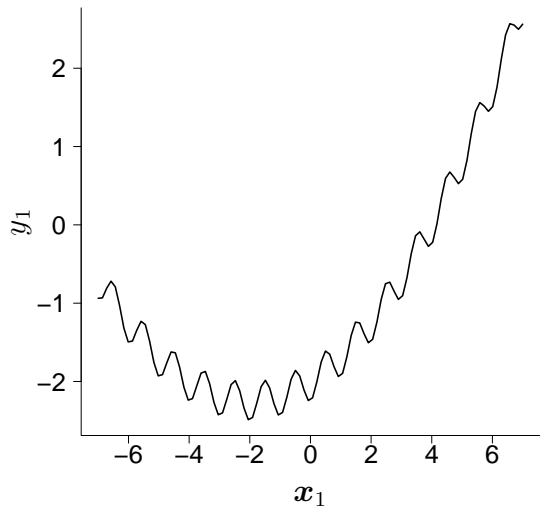
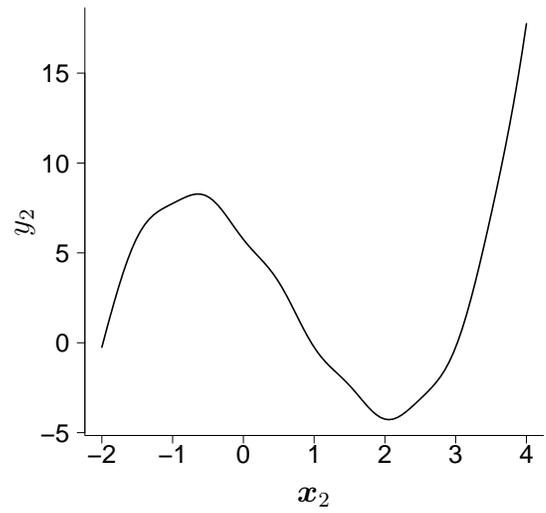
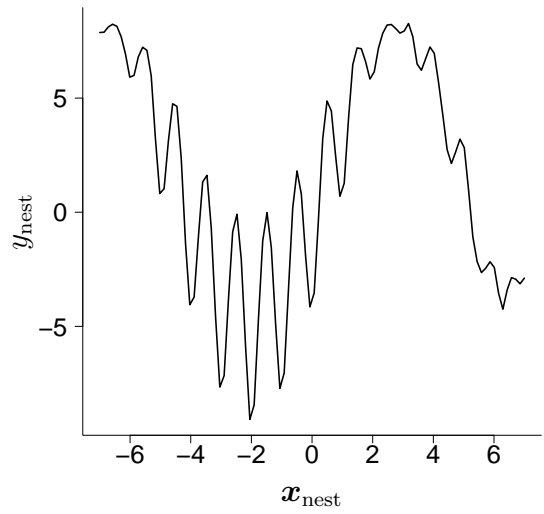
\caption{Analytical example: variations of the outputs $y_1$, $y_2$ and $y_{\text{nest}}$ of the codes 1, 2 and nested with respect to their input.
\label{fig0}
}
}

\subsubsection{Hydrodynamic example}

This example consists in the coupling of two computer codes. The objective is to determine the impact point of a conical projectile.

\hspace{0.cm}

The first code computes the drag coefficient of a cone divided by the height of the cone. Its inputs are the height and the half-angle of the cone, so the dimension of $\g{x}_1$ is 2.

\hspace{0.cm}

The second code computes the range of the ballistic trajectory of a cone. Its inputs are the output of the first code, associated with $\varphi_1$, and the initial velocity and angle of the ballistic trajectory of the cone, gathered in $\g{x}_2$. The dimension of $\g{x}_2$ is therefore 2.

\hspace{0.cm}

Figure \ref{fig1} illustrates the two codes inputs and outputs.

Figure \ref{fig2} shows the variations of the output with respect to each component of the input for each code. This figure enables to propose a basis of functions for the prior mean of the processes associated with the two codes.

For the first code the scatter plots highlight a linear variation with respect to $\PP{\g{x}_1}_1$ and a multiplicative inverse variation with respect to $\PP{\g{x}_1}_2$, so the proposed basis functions are: 

\Eq{\g{h}_1\PP{\g{x}_1}=\PP{1\ ,\ \PP{\g{x}_1}_1\ ,\ \dfrac{1}{\PP{\g{x}_1}_2}}^T.}

For the second code only a multiplicative inverse variation with respect to $y_1$ is evident, so the proposed basis functions are: 

\Eq{\g{h}_2\PP{\varphi_1,\g{x}_2}=\PP{1\ ,\  \dfrac{1}{\max\PP{\varphi_1,\varphi_{1_{\text{min}}}}}}^T.}
The denominator has a lower boundary $\varphi_{1_{\text{min}}}$ in order to avoid any inversion problem around zero. This boundary is small and set arbitrarily.

\renewcommand{\taillefigure}{0.37}

\BFig{
\Tabu{cc}{
\subfloat[Code 1: drag coefficient / height of the cone]{
\BPict{5.5}{8}
{
\centering
\pspolygon(0,3)(5,5)(5,1)
\psset{linestyle=dotted}
\psline(0,3)(5,3)
\psline{<->}(0,0.8)(5,0.8)
\psarc(0,3){2}{0}{21}
\rput(2.6,3.4){$\PP{\g{x}_1}_1$}
\rput(2.4,0.2){$\PP{\g{x}_1}_2$}
}
}
&
\psfrag{0.11}[c][c][1.2]{}
\psfrag{xlab}[c][c][1]{Horizontal distance}
\psfrag{ylab}[c][c][1]{Vertical distance}
\subfloat[Code 2: range of a ballistic trajectory]{
\IncGraph{\taillefigure}{\DossierRacine 24_CasTest_EvolutionVariables/2_Figures/1_TraceTrajectoire-1.eps}}

\psset{linecolor=lightgray}

\psset{linewidth=1.5mm}
\psset{arrowscale=1.5}
\psline{->}(-5.8,0.9)(-4.2,3)
\rput(-2.8,2.7){\textcolor{gray}{Initial velocity}}

\psset{linecolor=black}
\psset{linestyle=dotted}
\psset{linewidth=0.5mm}
\psline{}(-5.8,0.9)(-3.4,0.9)
\psarc(-5.8,0.9){1}{0}{55}
\rput(-4.3,1.5){$\PP{\g{x}_2}_2$}

\psline{<->}(-5.8,6)(-0.5,6)
\rput(-3.2,6.4){$y_2$}

\psline{<->}(-6.1,1)(-4.5,3.1)
\rput(-5.4,2.9){$\PP{\g{x}_2}_1$}
}
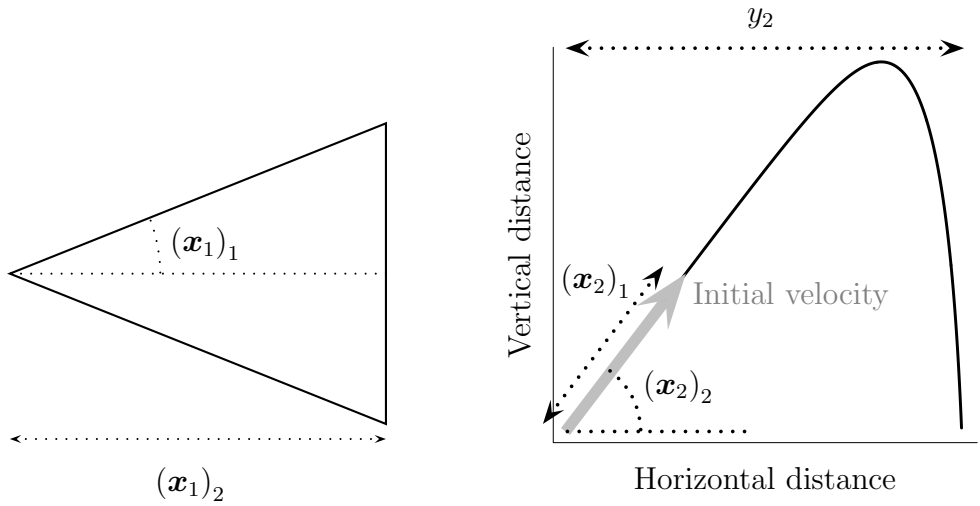
\caption{Hydrodynamic example: Inputs and outputs of the two codes.
\label{fig1}
}

}

\renewcommand{\taillefigure}{0.35}

\BFig{
\vspace{-0.7cm}
\Tabu{cc}{
\vspace{-0.7cm}
\psfrag{Titre}[c][c][1]{}
\psfrag{x}[c][c][1]{$\PP{\g{x}_1}_1$}
\psfrag{ylab}[c][c][1]{$y_1$}
\psfrag{x1}[c][c][0.9]{MC$100$}
\psfrag{x2}[c][c][0.9]{MC$1 000$}
\psfrag{x3}[c][c][0.9]{Linearized}
\subfloat[Code 1]{
\IncGraph{\taillefigure}{\DossierRacine 24_CasTest_EvolutionVariables/4_Figures/2_Code1-1.eps}}
&
\psfrag{Titre}[c][c][1]{}
\psfrag{x}[c][c][1]{$\PP{\g{x}_1}_2$}
\psfrag{ylab}[c][c][1]{$y_1$}
\psfrag{x1}[c][c][0.9]{MC$100$}
\psfrag{x2}[c][c][0.9]{MC$1 000$}
\psfrag{x3}[c][c][0.9]{Linearized}
\subfloat[Code 1]{
\IncGraph{\taillefigure}{\DossierRacine 24_CasTest_EvolutionVariables/4_Figures/2_Code1-2.eps}}
\\
\vspace{-0.7cm}
\psfrag{Titre}[c][c][1]{}
\psfrag{x}[c][c][1]{$\varphi_1= y_1$}
\psfrag{ylab}[c][c][1]{$y_2$}
\psfrag{x1}[c][c][0.9]{MC$100$}
\psfrag{x2}[c][c][0.9]{MC$1 000$}
\psfrag{x3}[c][c][0.9]{Linearized}
\subfloat[Code 2]{
\IncGraph{\taillefigure}{\DossierRacine 24_CasTest_EvolutionVariables/4_Figures/3_Code2-1.eps}}
&
\psfrag{Titre}[c][c][1]{}
\psfrag{x}[c][c][1]{$\PP{\g{x}_2}_1$}
\psfrag{ylab}[c][c][1]{$y_2$}
\psfrag{x1}[c][c][0.9]{MC$100$}
\psfrag{x2}[c][c][0.9]{MC$1 000$}
\psfrag{x3}[c][c][0.9]{Linearized}
\subfloat[Code 2]{
\IncGraph{\taillefigure}{\DossierRacine 24_CasTest_EvolutionVariables/4_Figures/3_Code2-2.eps}}
\\

\multicolumn{2}{c}{
\psfrag{Titre}[c][c][1]{}
\psfrag{x}[c][c][1]{$\PP{\g{x}_2}_2$}
\psfrag{ylab}[c][c][1]{$y_2$}
\psfrag{x1}[c][c][0.9]{MC$100$}
\psfrag{x2}[c][c][0.9]{MC$1 000$}
\psfrag{x3}[c][c][0.9]{Linearized}
\subfloat[Code 2]{
\IncGraph{\taillefigure}{\DossierRacine 24_CasTest_EvolutionVariables/4_Figures/3_Code2-3.eps}}
}
}
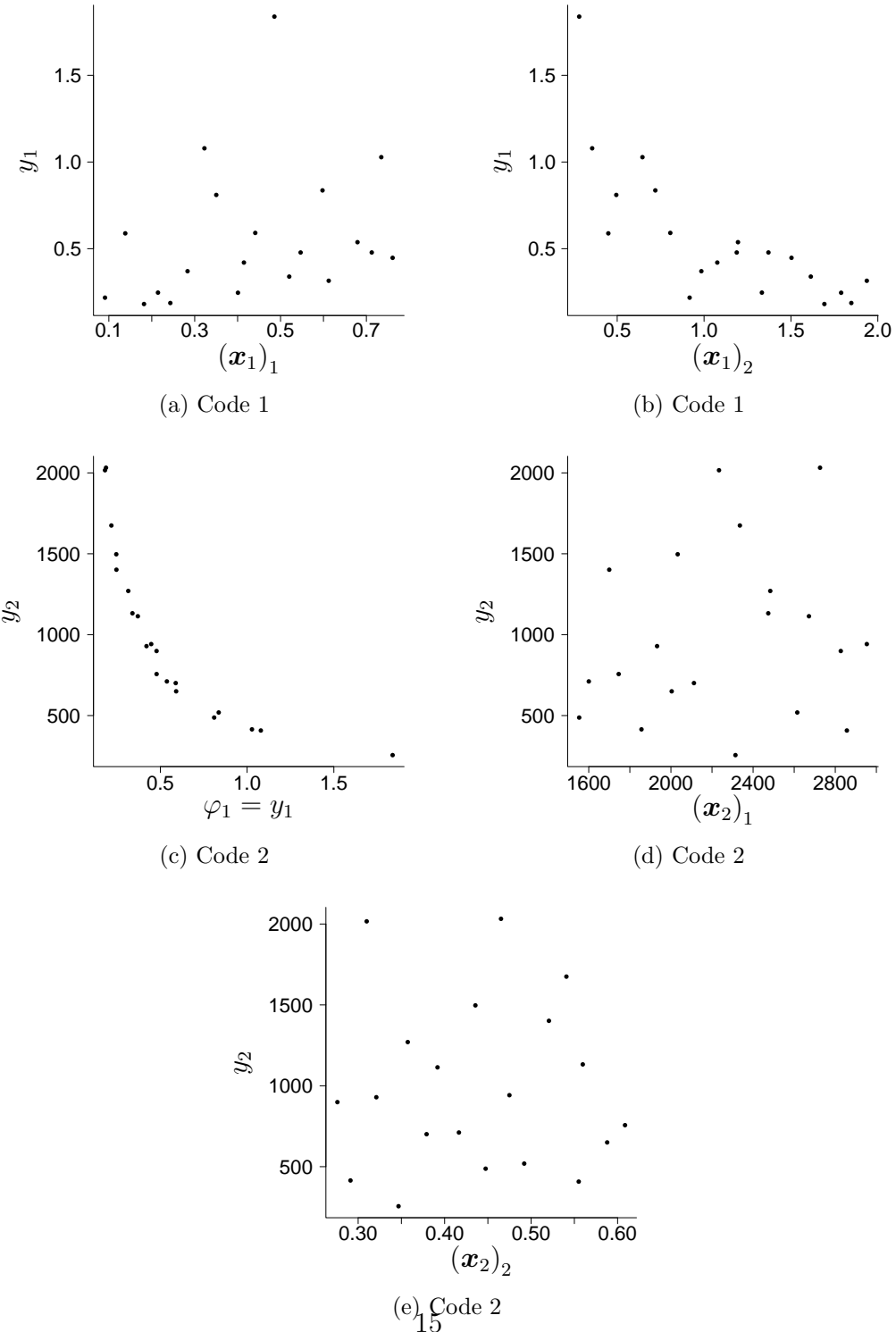
\caption{Hydrodynamic example: variation of the outputs $y_1$ and $y_2$ of the two codes with respect to the components of the inputs $\g{x}_1$ and $\g{x}_2$. The $20$ input points are drawn according to a maximin LHS design on $\mathbb{X}_1 \times\mathbb{X}_2$.
\label{fig2}
}
}

\subsection{Reference: "blind box" method}

In this method, the nested computer code is considered as a single computer code. Only the inputs $\g{x}_{\text{nest}}$ and the output $y_{\text{nest}}$ are taken into account. The intermediary information $\varphi_1$ is not considered. A Gaussian process regression of this single computer code is done.

Only the chained I-optimal sequential design could be applied in this framework, the other proposed sequential design requiring to consider the partial information.

\subsection{Choice of the covariance functions and estimation of their hyperparameters}

In the analytical example the covariance functions are Gaussian. This implies that the sample paths of the Gaussian processes associated with the codes are infinitely differentiable functions. This enables to apply Proposition \ref{prop2} and Proposition \ref{prop3} to this example.

In the hydrodynamic example the covariance functions are Mat\'ern $\frac{5}{2}$, which implies that the sample paths of the Gaussian processes associated with the codes are mean square one time continuously differentiable functions (see \cite{Rasmussen2006}). This enables to perform the linearization of Proposition \ref{prop3}.

In both cases the covariance functions include a non-zero nugget term (see \cite{Gramacy2012Stat} for further details).

\hspace{0cm}

The hyperparameters of the covariance functions are estimated for each set of observations, including the sequential designs. They are estimated by maximizing the Leave-One-Out log predictive probability (see \cite{Rasmussen2006}, chapter 5, and \cite{Bachoc2013}).

\subsection{Comparison between the analytical and the linearized method}

Figure \ref{fig3} illustrates the convergence of the two first statistical moments estimated with the Monte Carlo (see Proposition \ref{prop1}) and the linearized methods (see Proposition \ref{prop3}) towards their real values calculated with the analytical method described in Proposition \ref{prop2}. 

Both methods converge when the uncertainty of the first code predictor decreases.
It can be seen that the linearized method is a very good compromise between computation time and accuracy compared to the Monte Carlo method.

\renewcommand{\taillefigure}{0.37}

\BFig{
\vspace{-0.7cm}
\Tabu{cc}{

\psfrag{xlab}[c][c][1]{}
\psfrag{ylab}[c][c][1]{\textit{Time (s)}}
\psfrag{x1}[c][c][0.9]{MC$100$}
\psfrag{x2}[c][c][0.9]{MC$1 000$}
\psfrag{x3}[c][c][0.9]{Linearized}

\subfloat[Calculation time]{
\IncGraph{\taillefigure}{\DossierRacine 13_ConvergenceMonteCarloAnalytique/16_Figures/0_TempsCalcul-1.eps}}
&

\hspace{-0.8cm}
\psfrag{Titre}[c][c][1.2]{}
\psfrag{xlab}[c][c][1]{\textit{$\GPCondEcartType{1}$}}
\psfrag{ylab}[c][c][1]{\textit{Relative error of the estimation}}
\psfrag{x1}[l][l][0.8]{MC $100$}
\psfrag{x2}[l][l][0.8]{MC $1 000$}
\psfrag{x3}[l][l][0.8]{Linearized}
\subfloat[First moment]{
\IncGraph{\taillefigure}{\DossierRacine 13_ConvergenceMonteCarloAnalytique/16_Figures/1_M1-2.eps}}
\hspace{1cm}	
\\
\multicolumn{2}{c}{
\psfrag{Titre}[c][c][1.2]{}
\psfrag{xlab}[c][c][1]{\textit{$\GPCondEcartType{1}$}}
\psfrag{ylab}[c][c][1]{\textit{Relative error of the estimation}}
\psfrag{x1}[l][l][0.8]{MC $100$}
\psfrag{x2}[l][l][0.8]{MC $1 000$}
\psfrag{x3}[l][l][0.8]{Linearized}
\subfloat[Second moment]{
\IncGraph{\taillefigure}{\DossierRacine 13_ConvergenceMonteCarloAnalytique/16_Figures/2_M2-2.eps}}
}
\\
}
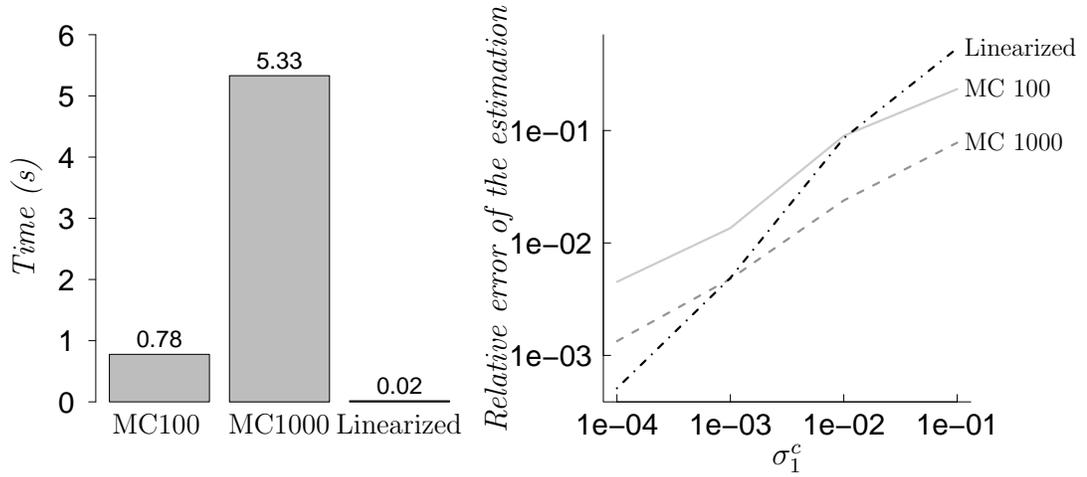
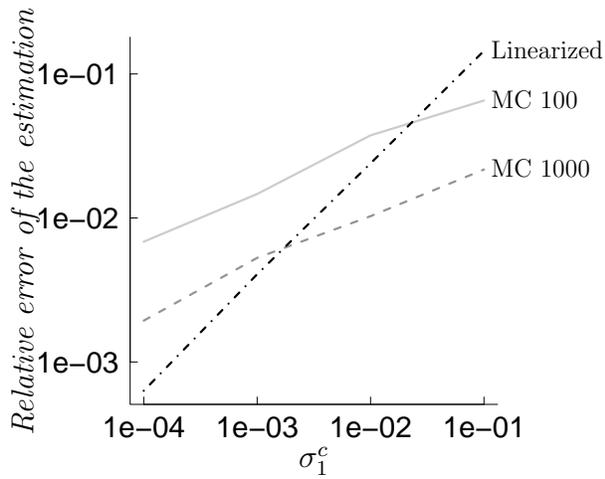
\caption{Comparison of the linearized (Proposition \ref{prop3}) and Monte-Carlo (Proposition \ref{prop1}) methods in terms of computation time and accuracy for the evaluation of the two first moments of the process $\widehat{y}^c_\text{nest}$. The Monte Carlo method is run with $100$ and $1 000$ points to compute the one-dimensional integral with a Gaussian measure. The Monte Carlo draws are repeated $50$ times and the curves correspond to the median of these repetitions. \newline The real values are computed with the analytical method (Proposition \ref{prop2}). The covariance functions are Gaussian. 
The predictor of the first code is of the form $y_1^c = \mu_1^c + \sigma_1^c u$ with $u \sim \Normale{0}{1}$, $\sigma_1^c \in \Ac{10^{-4},10^{-3},10^{-2},10^{-1}}$ and for each value of $\sigma_1^c$, $100$ values of $\mu_1^c $ on a grid on $\Cr{-2,4}$ are considered.
The predictor of the second code is build using $20$ input observation points drawn on a grid on $\Cr{-2,4}$ for the second code of the analytical example.
\label{fig3}
}
}

\subsection{Definition of the performance criterion of the predictor mean}

A set of validation observations if available. Let $\g{x}_\text{nest}^{(1)} \dots \g{x}_\text{nest}^{(N_\text{nest})}$ be $N_\text{nest}$ elements of $\mathbb{X}_\text{nest}$. \\Denoting by 
$y_\text{nest}\PP{\g{x}_\text{nest}^{(1)}} \dots y_\text{nest}\PP{\g{x}_\text{nest}^{(N_\text{nest})}}$ the evaluations of the nested code in these points, the performance criterion of the nested predictor mean, also called error on the mean can be defined as: 

\Eq{
\text{Error on the mean}= \dfrac{
\sum \limits_{i=1}^{N_\text{nest}} \PP{y_\text{nest}\PP{\g{x}_\text{nest}^{(i)}} - \Mean{\widehat{y}^c_\text{nest}\PP{\g{x}_\text{nest}^{(i)}}} }^2
}{
\sum \limits_{i=1}^{N_\text{nest}} \PP{y_\text{nest}\PP{\g{x}_\text{nest}^{(i)}} - \dfrac{1}{N_\text{nest}}\sum \limits_{j=1}^{N_\text{nest}} y_\text{nest}\PP{\g{x}_\text{nest}^{(j)}} }^2
}.
}

\subsection{Comparison between the blind box and the linearized methods}

Figure \ref{fig4} shows that the linearized method enables to better take into account the non-stationarity of the variations of the nested code output. On the contrary, in the blind box method the magnitude of the prediction interval is the same across the input domain and depends only on the distance to the observation points. The prediction interval is too big in the area with small variations and too small in the area with larger variations.

\hspace{0cm}

\BFig{
\Tabu{cc}{
\psfrag{Titre}[c][c][1]{}
\psfrag{xlab}[c][c][1]{$\g{x}_\text{nest}$}
\psfrag{ylab}[c][c][1]{$y_\text{nest}$}
\subfloat[Linearized method]{
\IncGraph{\taillefigure}{\DossierRacine 25_ExemplePoly_TraceFonctions/2_Figures/1_TracePredicteurs20-1.eps}}
&
\psfrag{Titre}[c][c][1]{}
\psfrag{xlab}[c][c][1]{$\g{x}_\text{nest}$}
\psfrag{ylab}[c][c][1]{$y_\text{nest}$}
\subfloat[Blind box method]{
\IncGraph{\taillefigure}{\DossierRacine 25_ExemplePoly_TraceFonctions/2_Figures/1_TracePredicteurs20-2.eps}}
\\

}
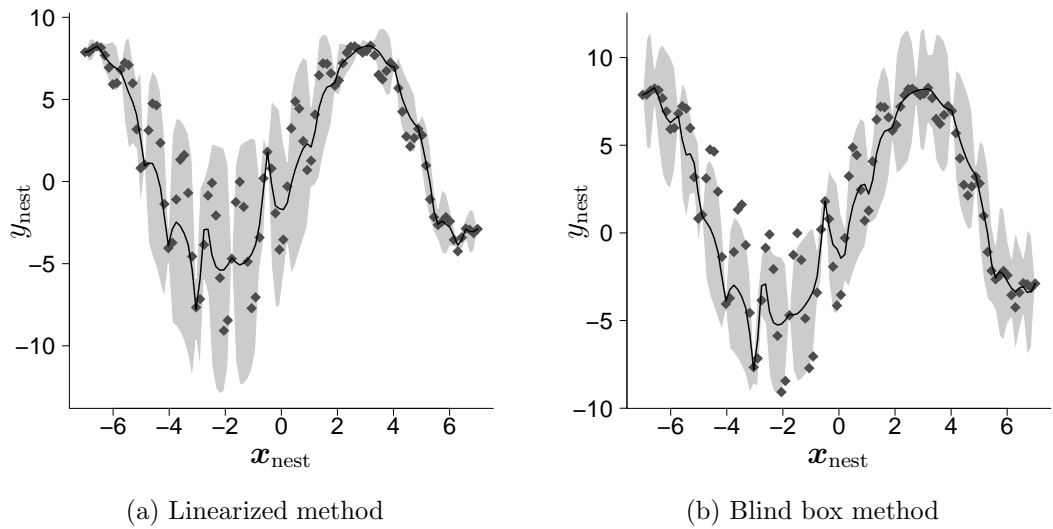
\caption{Analytical example: Predictors of the nested code obtained with the linearized and the blind box methods. The set of 20 observations is drawn according to a maximin LHS on $\mathbb{X}_1$. Actual values shown by dots, the mean of prediction by a line and the 95\% prediction interval of prediction by a grey area.
\label{fig4}
}
}

Figure \ref{fig5} shows the similar accuracies of the prediction mean computed with the analytical and linearized methods proposed in Proposition \ref{prop2} and Proposition \ref{prop3}.

For both examples, the precision of the prediction mean is better with the linearized method than with the blind box method, showing the interest of taking into account the intermediary information.

\renewcommand{\taillefigure}{0.44}

\BFig{
\vspace{-0.7cm}
\Tabu{cc}{
\hspace{-0.5cm}
\psfrag{Titre}[c][c][1.2]{}
\psfrag{xlab}[c][c][1]{\textit{Number of chained evaluations}}
\psfrag{ylab}[c][c][1]{\textit{Error on the mean}}
\psfrag{BB}[l][c][0.8]{Blind box}
\psfrag{Lin}[l][c][0.8]{Linearized}
\psfrag{AN}[l][c][0.8]{Analytical}
\subfloat[Analytical: Gaussian covariance]{\IncGraph{\taillefigure}{\DossierRacine 14_ComparaisonMethodes_ExemplePoly_CovGauss/7_Figures/1_MSE-1.eps}}
&
\hspace{-0.7cm}
\psfrag{Titre}[c][c][1.2]{}
\psfrag{xlab}[c][c][1]{\textit{Number of chained evaluations}}
\psfrag{ylab}[c][c][1]{\textit{Error on the mean}}
\psfrag{BB}[l][c][0.8]{Blind box}
\psfrag{Lin}[l][c][0.8]{Linearized}
\subfloat[Hydrodynamic example: Mat\'ern $\dfrac{5}{2}$ covariance]{\IncGraph{\taillefigure}{\DossierRacine 20_ComparaisonMethodes_CasTest_CovMatern/7_Figures/1_MSE-1.eps}}
}
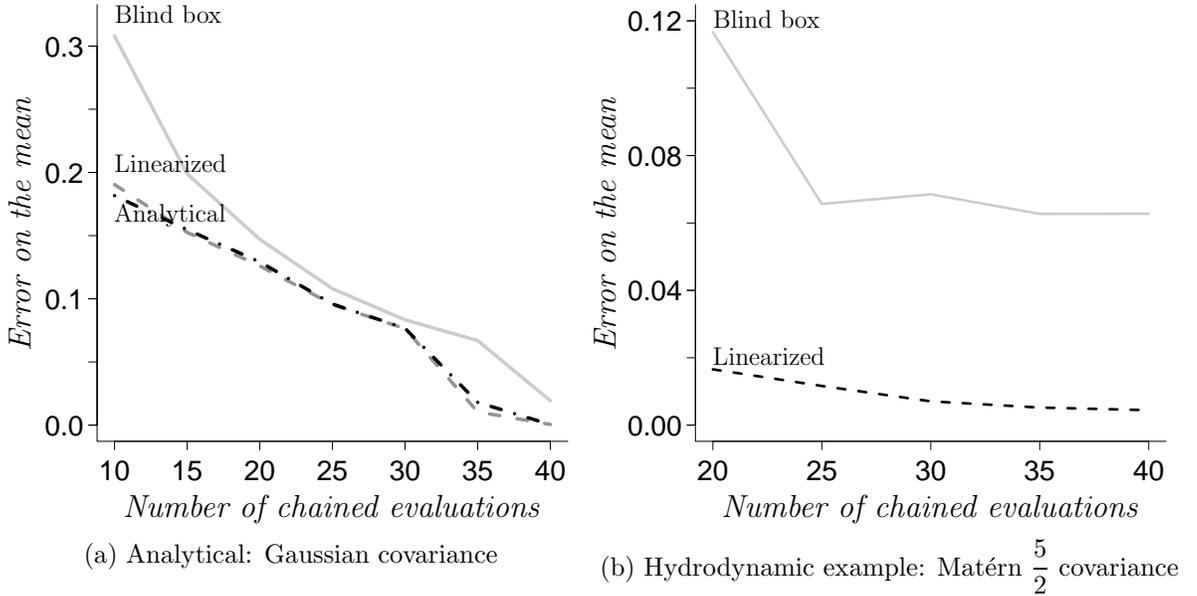
\caption{Comparison of the prediction mean accuracy for the blind box and the linearized (Proposition \ref{prop3}) methods, and, in case of a Gaussian covariance function, the analytical method (Proposition \ref{prop2}). The curves correspond to the median of $50$ draws of maximin LHS designs on $\mathbb{X}_1 \times \mathbb{X}_2$ of increasing size.
\label{fig5}
}
}

\subsection{Performances of the sequential designs with identical computational costs}

Figure \ref{fig6} shows the relevance of the proposed sequential designs for improving the prediction mean of the linearized nested predictor, compared to the maximin LHS design on $\mathbb{X}_{\text{nest}}$. 

In the analytical example, the best I-optimal sequential design enables to obtain the most accurate prediction mean at a given computational cost. In the hydrodynamic example, the different sequential designs give similar results, except for the first new observation points added, where the best I-optimal is better.

\renewcommand{\taillefigure}{0.45}
\BFig{
\hspace{-3cm}
\Tabu{c}{
\psfrag{Titre}[c][c][1.2]{}
\psfrag{xlab}[c][c][1]{\textit{Computational cost}}
\psfrag{ylab}[c][c][1]{\textit{Error on the mean}}
\psfrag{x1}[l][c][0.8]{LHS}
\psfrag{x2}[l][c][0.8]{Chained I-optimal}
\psfrag{x3}[l][c][0.8]{Best I-optimal}
\subfloat[Analytical example]{\IncGraph{\taillefigure}{\DossierRacine 27_ComparaisonMethodes_ExemplePoly/14_Figures/1_MSE-1.eps}}
\\[-30pt]
\psfrag{Titre}[c][c][1.2]{}
\psfrag{xlab}[c][c][1]{\textit{Computational cost}}
\psfrag{ylab}[c][c][1]{\textit{Error on the mean}}
\psfrag{x1}[l][c][0.8]{LHS}
\psfrag{x2}[l][c][0.8]{Chained I-optimal}
\psfrag{x3}[l][c][0.8]{Best I-optimal}
\subfloat[Hydrodynamic example]{\IncGraph{\taillefigure}{\DossierRacine 26_ComparaisonMethodes_CasTest/14_Figures/1_MSE-1.eps}}
\\
}
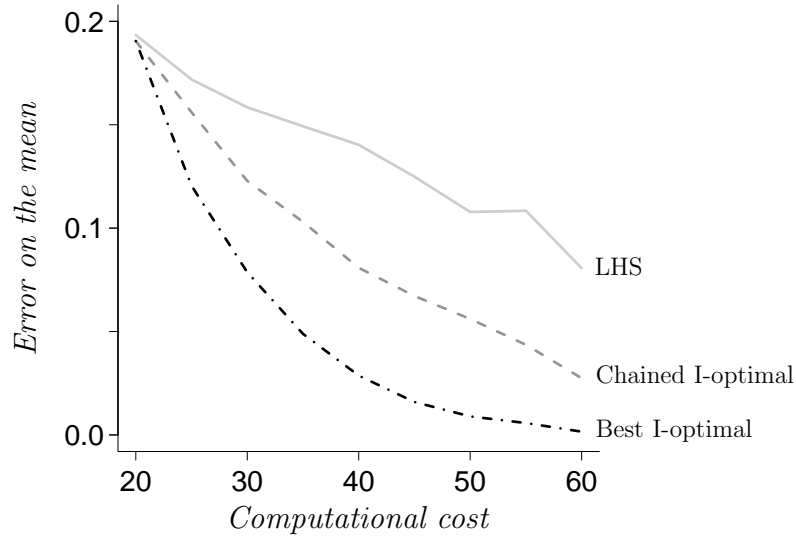
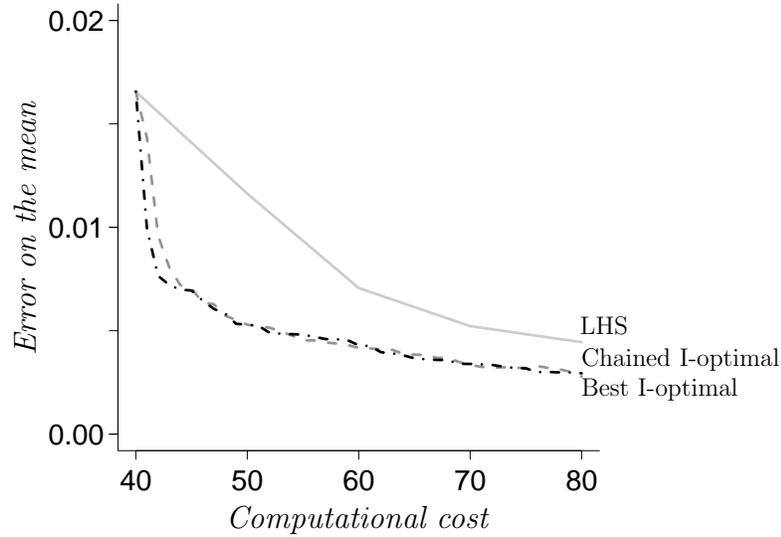
\caption{Comparison of the linearized predictor mean precision with the maximin LHS design on $\mathbb{X}_\text{nest}$ and the sequential designs applied to the two examples. In the hydrodynamic example, the two curves representing the sequential designs are almost superimposed. The initial designs are the same for the three curves, with a size of $10$ points for the analytical example and $20$ points for the hydrodynamical example. The draw of the chained maximin LHS designs is repeated $50$ times and the curves present the median of the associated results. The costs of the two codes are assumed to be the same.
\label{fig6}
}
}

\hspace{0cm}

In both examples the new observation points are mostly added on the first code, as shown in figure \ref{fig7}.
It seems that the uncertainty propagated from the first code into the second code is predominant at the beginning. The best I-optimal sequential design aims therefore to reduce this uncertainty by first adding new observation points on the first code. Then new observations points can be added on both codes.

\BFig{
\vspace{-0.7cm}
\Tabu{cc}{
\hspace{-1cm}
\psfrag{Titre}[c][c][1.2]{}
\psfrag{xlab}[c][c][1]{\textit{Computational cost}}
\psfrag{ylab}[c][c][1]{\textit{Number of code evaluations}}
\psfrag{x1}[l][l][0.8]{Code 1}
\psfrag{x2}[l][l][0.8]{Code 2}
\subfloat[Analytical example]{\IncGraph{\taillefigure}{\DossierRacine 17_ComparaisonAmeliorationSequentielle_ExemplePoly_CovMatern/10_Figures/2_AppelsCodes-1.eps}}

&
\hspace{-0.8cm}
\psfrag{Titre}[c][c][1.2]{}
\psfrag{xlab}[c][c][1]{\textit{Computational cost}}
\psfrag{ylab}[c][c][1]{\textit{Number of code evaluations}}
\psfrag{x1}[l][l][0.8]{Code 1}
\psfrag{x2}[l][l][0.8]{Code 2}
\subfloat[Hydrodynamic example]{\IncGraph{\taillefigure}{\DossierRacine 21_AmeliorationSequentielle_CasTest_Cout11/10_Figures/2_AppelsCodes-1.eps}}
\\
}
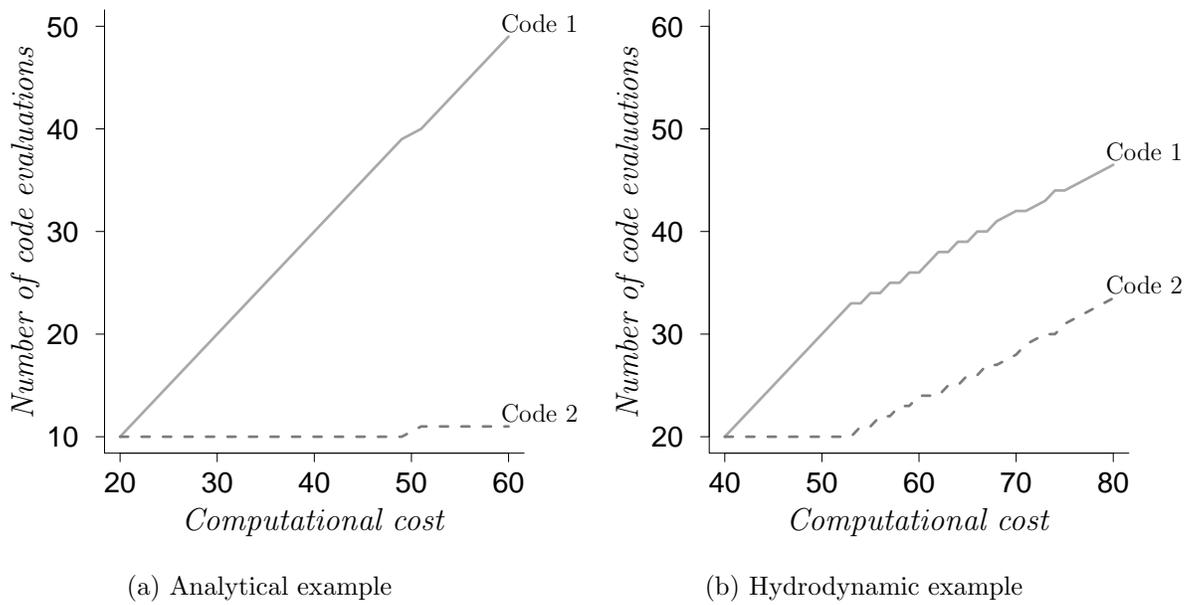
\caption{Comparison of the number of evaluations of each code in case of a sequential best I-optimal design applied to both examples. The curves correspond to the median of $50$ draws of the initial design. The costs of the two codes are assumed to be the same.
\label{fig7}
}
}

\subsection{Performances of the sequential designs with different computational costs}

Figure \ref{fig8} shows the prediction mean accuracy with a best I-optimal sequential design when the costs of the two codes are different. It can be seen that at a given total computational cost the accuracy of prediction is better when the cost of the first code is lower. In other words the prediction mean accuracy is better at a given computational budget when more observation points can be added to the first code for the same computational budget.
These results are consistent with those of figure \ref{fig7}.

\renewcommand{\taillefigure}{0.43}
\BFig{
\vspace{-0.7cm}
\Tabu{cc}{

\psfrag{Titre}[c][c][1.2]{}
\psfrag{xlab}[c][c][1]{\textit{Computational cost}}
\psfrag{ylab}[c][c][1]{\textit{Error on the mean}}
\psfrag{x1}[l][c][0.8]{ 1:2}
\psfrag{x2}[l][c][0.8]{ 2:1}	
\subfloat[Analytical example]{\IncGraph{\taillefigure}{\DossierRacine 19_AmeliorationSequentielle_ExemplePoly_Cout21/12_Figures/1_MSE-1.eps}}
&
\hspace*{-0.8cm}
\psfrag{Titre}[c][c][1.2]{}
\psfrag{xlab}[c][c][1]{\textit{Computational cost}}
\psfrag{ylab}[c][c][1]{\textit{Error on the mean}}
\psfrag{x1}[l][c][0.8]{1:2}
\psfrag{x2}[l][c][0.8]{2:1}
\subfloat[Hydrodynamic example]{\IncGraph{\taillefigure}{\DossierRacine 23_AmeliorationSequentielle_CasTest_Cout21/12_Figures/1_MSE-1.eps}}
\\

}
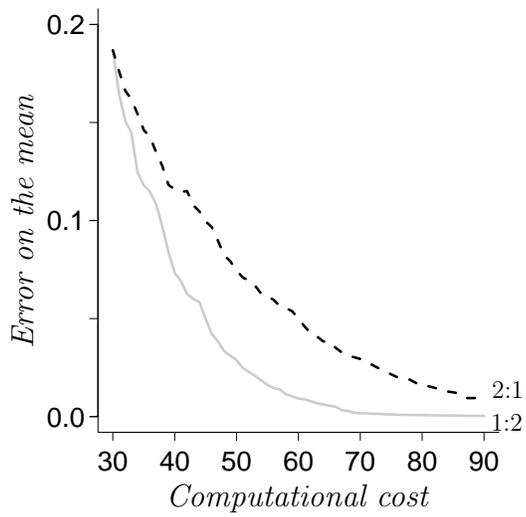
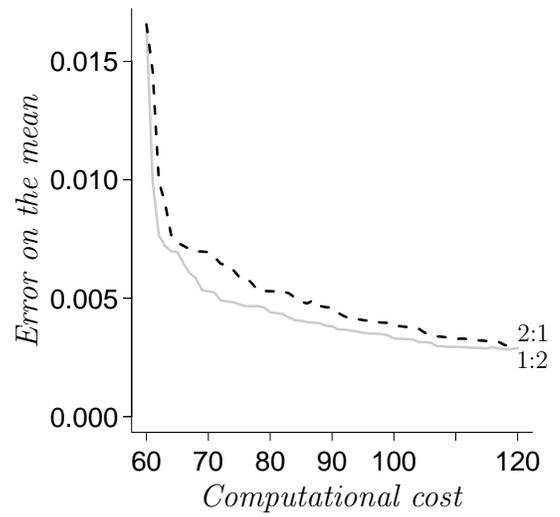
\caption{Performances of the best I-optimal sequential design in terms of prediction mean accuracy with different computational costs for the two codes. 1:2 $\leftrightarrow$ cost 1 for code 1 and 2 for code 2, 2:1 $\leftrightarrow$ cost 2 for code 1 and 1 for code 2. The curves correspond to the median of $50$ draws of the initial maximin LHS design on $\mathbb{X}_\text{nest}$. The initial designs are the same for the two curves corresponding to each example and contain 15 observations and 30 observations on both codes for the analytical and the hydrodynamical example.
\label{fig8}
}
}

\section{Conclusions and future work}

In this paper the Gaussian process formalism is adapted to the case of two nested computer codes.\\
Two methods to evaluate quickly the mean and variance of the nested code predictor have been proposed. The first one, called "analytical" computes the exact value of the two first moments of the predictor. But it cannot be applied to the coupling of more than two codes. The second one, called "linearized", enables to obtain a Gaussian predictor of the two nested codes, with mean and variance that can be instantly computed. The approach could be generalized to the coupling of more than two codes.\\
Both proposed methods take into account the intermediary information, that means the output of the first code. A comparison to the reference method, called "blind box", is made. In this method a Gaussian process regression of the block of the two codes is made without considering the intermediary observations. The numerical examples illustrate the interest of taking into account the intermediary information in terms of prediction mean accuracy. 

\vspace*{0.2cm}

Moreover, two sequential designs are proposed in order to improve the prediction accuracy of the nested predictor. The first one, the "chained" I-optimal sequential design, corresponds to the case when the two codes cannot be launched separately. The second one, the "best" I-optimal sequential design, allows to choose on which of the two codes to add a new observation point and to take into account the different computational costs of the two codes.\\
The numerical applications show the interest of the sequential designs compared to a space-filling design (maximin LHS). Furthermore, they illustrate the advantage, in terms of prediction mean accuracy, of choosing on which code to add a new observation point compared to simply adding new observation points of the nested code. The results obtained show an amplification of the uncertainties in the chain of codes, leading to the addition of observation points on the first code firstly in the best I-optimal sequential design. It can be assumed that this should be similar with the coupling of more than two codes. In other words, the uncertainty of the beginning of the chain should be reduced as a priority.

\vspace*{0.5cm}

This paper has been focused on the case of two nested codes with a scalar intermediary variable. Considering the case of a functional intermediary variable seems promising for future work.

\newpage
\section*{Appendix}

\subsection*{Proof of Proposition \ref{prop1}}
According to Eq \eqref{predictGP}: 
\[\widehat{y}^c_i\PP{\g{x_i}}=\mu_i^{c}\PP{\g{x_i}}+\sigma_i^c\PP{\g{x_i}}\xi_i,\quad    \xi_i\sim\mathcal{N}(0,1),\quad i\in\Ac{1,2},\]

\noindent where $\xi_1$ and $\xi_2$ are independent according to the independence of the initial processes $\widehat{y}_1$ and $\widehat{y}_2$.

Therefore the process modeling the nested code can be written:
\[
\Tab{rl}{
\widehat{y}_{\text{nest}}^c(\g{x}_1,\g{x}_2)=
&
\widehat{y}^c_2(\widehat{y}^c_1(\g{x}_1),\g{x}_2)
\\[5pt]
=&
\mu_2^{c}\PP{\mu_1^{c}\PP{\g{x_1}}+\sigma_1^c\PP{\g{x_1}}\xi_1,\g{x_2}}
+\sigma_2^c\PP{\mu_1^{c}\PP{\g{x_1}}+\sigma_1^c\PP{\g{x_1}}\xi_1,\g{x_2}}\xi_2 
}
\]

Given the independence of $\xi_1$ and $\xi_2$ and the fact that $\mathbb{E}\PP{\xi_2}=0$, it can be inferred that the first moment of $\widehat{y}^c_\text{nest}$ can be written:
\[
\mathbb{E}\PP{\widehat{y}_{\text{nest}}^c(\g{x}_1,\g{x}_2)}
=\mathbb{E}\PP{\mu_2^{c}\PP{\mu_1^{c}\PP{\g{x_1}}+\sigma_1^c\PP{\g{x_1}}\xi_1,\g{x_2}}}
\]

By noting that: 
\BI{
\item
$
\Tab{rl}{
\PP{\widehat{y}_{\text{nest}}^c(\g{x}_1,\g{x}_2)}^2=
&
\PP{\widehat{y}^c_2(\widehat{y}^c_1(\g{x}_1),\g{x}_2)}^2
\\[5pt]
=&
\PP{\mu_2^{c}\PP{\mu_1^{c}\PP{\g{x_1}}+\sigma_1^c\PP{\g{x_1}}\xi_1,\g{x_2}}
+\sigma_2^c\PP{\mu_1^{c}\PP{\g{x_1}}+\sigma_1^c\PP{\g{x_1}}\xi_1,\g{x_2}}\xi_2 }
\\[5pt]
=&
\PP{\mu_2^{c}\PP{\mu_1^{c}\PP{\g{x_1}}+\sigma_1^c\PP{\g{x_1}}\xi_1,\g{x_2}}}^2
+\PP{\sigma_2^c\PP{\mu_1^{c}\PP{\g{x_1}}+\sigma_1^c\PP{\g{x_1}}\xi_1,\g{x_2}} }^2\xi_2^2
\\[5pt]
&
+2 \mu_2^{c}\PP{\mu_1^{c}\PP{\g{x_1}}+\sigma_1^c\PP{\g{x_1}}\xi_1,\g{x_2}}\sigma_2^c\PP{\mu_1^{c}\PP{\g{x_1}}+\sigma_1^c\PP{\g{x_1}}\xi_1,\g{x_2}}\xi_2
}
$

\item $\xi_1$ and $\xi_2$ are independent,

\item $\mathbb{E}\PP{\xi_2}=0$ and $\mathbb{E}\PP{\xi_2^2}=1$,
}
the second moment of $\widehat{y}^c_\text{nest}$ can be written:
\[
\mathbb{E}\PP{\PP{\widehat{y}^c_2(\widehat{y}^c_1(\g{x}_1),\g{x}_2)}^2}
=\mathbb{E}\Cr{
\Tab{l}
{
\PP{\mu_2^{c}\PP{\mu_1^{c}\PP{\g{x_1}}+\sigma_1^c\PP{\g{x_1}}\xi_1,\g{x_2}}}^2
\\[5pt]
+\PP{\sigma_2^c\PP{\mu_1^{c}\PP{\g{x_1}}+\sigma_1^c\PP{\g{x_1}}\xi_1,\g{x_2}} }^2
}
}
\]

\vspace{1cm}

\subsection*{Proof of Proposition \ref{prop2}}

If $x \sim \mathcal{N}(\mu,\sigma^2)$ and $f\PP{x,a,b,c}=x^c\exp\PP{a x + b x^2}$ then the mean of $f\PP{x,a,b,c}$ is defined as:
\[
\Mean{f\PP{x,a,b,c}}=\exp\PP{ -\dfrac{1}{2\sigma^2}\PP{
\frac{ \PP{\sigma^2 a + \mu}^2 }{2\sigma^2 b -1} + \mu^2	} 
}
\Mean{x_f^c}
\]

where 
$
x_f \sim 
\Normale{\dfrac{\sigma^2 a+\mu}{1-2b\sigma^2}}
{\dfrac{\sigma^2}{1-2b\sigma^2}},
$
under the condition that $1-2b\sigma^2>0$.

Given that the moments of a Gaussian variable can be calculated analytically, $\Mean{x_g^c}$ and therefore $\Mean{f\PP{x,a,b,c}}$ can be computed analytically.

So we have shown that if $x \sim \mathcal{N}(\mu,\sigma^2)$, and $f\PP{x,a,b,c}=x^c\exp\PP{a x + b x^2}$ then, under the integrability condition $1-2b\sigma^2>0$, the mean of $f\PP{x,a,b,c}$ can be calculated analytically. 

\vspace*{0.8cm}

\subsubsection*{First moment}

In the framework of Universal Kriging, the conditional mean function of the process modeling the second code can be written:
\[
\Tab{rl}{
\mu_2^{c}\PP{\EntreeCodeDeux{}} =& 
\g{h_2}\PP{\EntreeCodeDeux{}}^T \g{v}_h
+
C_2\PP{\EntreeCodeDeux{},\ObsEntreeCodeDeux{}}\g{v}_c
\\[10 pt]
=&
\sum \limits_{i=1}^{M_2} \PP{\g{h_2}\PP{\EntreeCodeDeux{}}}_i \PP{\g{v}_h}_i
+
\sum \limits_{i=1}^{N_1} C_2\PP{\EntreeCodeDeux{},\EntreeCodeDeux{(i)}}\PP{\g{v}_c}_i
\\[10 pt]
=&
(1)
+
(2)
}
\]
\noindent where $\g{v}_h \in \R^{M_2}$ and $\g{v}_c \in \R^{N_1}$ and $\varphi_1 \sim \Normale{\mu_1^c}{\PP{\sigma_1^c}^2}$.

\vspace{0.5cm}

According to the assumptions of Proposition \ref{prop2} the mean basis functions $\g{h}_2$ can be written: 
\[
(\g{h}_2(\EntreeCodeDeux{})_i=m_i(\g{x}_2)\ f\PP{\varphi_1,\PP{\g{\alpha}_i}_3,\PP{\g{\alpha}_i}_1,\PP{\g{\alpha}_i}_2},
\]
with $m_i$ deterministic functions.

In the same way, the covariance function $C_2$ can be written: 
\[
C_2\PP{\EntreeCodeDeux{},\EntreeCodeDeux{'}} = 
\sigma_2^2 \dfrac{1}{l_{\varphi_1}}\ k^{(2 n_{\varphi_1})}\PP{\dfrac{\varphi_1-\varphi'_1}{l_{\varphi_1}}} 
\prod \limits_{i=1}^{d_2} 
 \PP{ \dfrac{1}{l_i}\ k^{(2 n_i)}\PP{\dfrac{\PP{\g{x}_2}_i-\PP{\g{x}'_2}_i}{l_{i}}} }
,
\]

with $k: x \mapsto \exp\PP{-x^2/2}$, $n_{\varphi_1}$ and $n_i$ positive integers and $k^{(n)}$ denoting the n-th derivative of function $k$.
So, we can written that: 
\[
C_2\PP{\EntreeCodeDeux{},\EntreeCodeDeux{'}} = 
\sigma_2^2 \sum \limits_{j=1}^{n_{\varphi_1}} a_j f\PP{\varphi_1-\varphi'_1,0 , \dfrac{-1}{2l^2_{1}},2j}  l\PP{\g{x}_2-\g{x}'_2}\ 
,
\]
where $l$ is a deterministic function defined according to the previous equation and $a_j$ real numbers.

\vspace{0.5cm}
So the terms $(1)$ and $(2)$ of the previous equation can be written:
\[
\Tab{rl}{
(1) =
& 
\sum \limits_{i=1}^{M_2}  f\PP{\varphi_1,\PP{\g{\alpha}_i}_3,\PP{\g{\alpha}_i}_1,\PP{\g{\alpha}_i}_2}\ m_i(\g{x}_2)  \PP{\g{v}_h}_i
\\[5 pt]
}
\]

\[
\Tab{rl}{
(2) =
& 
\sum \limits_{i=1}^{N_1} \sigma_2^2 l\PP{\g{x}_2-\g{x}^{(i)}_2}\PP{\g{v}_c}_i \sum \limits_{j=1}^{n_{\varphi_1}} a_j f\PP{\varphi_1-\varphi^{(i)}_1,0 , \dfrac{-1}{2l^2_{1}},2j}\  
\\[5 pt]
}
\]

According to the fact that $m_i$ and $l$ are deterministic functions, $\g{v}_h$, $\g{v}_c$, $\g{x}_2^{(i)}$ and $\g{x}_2$ deterministic vectors, and $\varphi^{(i)}$ and $a_j$ deterministic real numbers, then: 
\[
\Tab{rl}{
\Mean{(1)} =
& 
\sum \limits_{i=1}^{M_2} \Mean{f\PP{\varphi_1,\PP{\g{\alpha}_i}_3,\PP{\g{\alpha}_i}_1,\PP{\g{\alpha}_i}_2}}\ m_i(\g{x}_2) \PP{\g{v}_h}_i
\\[5 pt]
}
\]

\[
\Tab{rl}{
\Mean{(2)} =
& 
\sum \limits_{i=1}^{N_1} \sigma_2^2 l\PP{\g{x}_2-\g{x}^{(i)}_2}\PP{\g{v}_c}_i \sum \limits_{j=1}^{n_{\varphi_1}} a_j \Mean{f\PP{\varphi_1-\varphi^{(i)}_1,0 , \dfrac{-1}{2l^2_{1}},2j}}\  
\\[5 pt]
}
\]
The means $\Mean{(1)}$ and $\Mean{(2)}$ can therefore be calculated analytically, and consequently, the mean $\Mean{\mu_2^{c}\PP{\EntreeCodeDeux{}} }$ can be calculated analytically.

\vspace*{0.8cm}
\subsubsection*{Second moment}

In the framework of Universal Kriging, it can be written that:
\[
\Tab{c}
{
\PP{\mu_2^{c}\PP{\EntreeCodeDeux{}}}^2+\PP{\sigma_2^{c}\PP{\EntreeCodeDeux{}}}^2=\sigma^2_2 

\\[5 pt]
+\underbrace{\g{h_2}\PP{\EntreeCodeDeux{}}^T \g{A}_h \g{h_2}\PP{\EntreeCodeDeux{}}}_\text{(1)} 
\\[20 pt]
+\underbrace{C_2\PP{\EntreeCodeDeux{},\ObsEntreeCodeDeux{}} \g{A}_c\ C_2\PP{\ObsEntreeCodeDeux{},\EntreeCodeDeux{}}}_\text{(2)} 
\\[20pt]
+\underbrace{C_2\PP{\EntreeCodeDeux{},\ObsEntreeCodeDeux{}} \g{A}_{ch}\ \g{h_2}\PP{\EntreeCodeDeux{}}}_\text{(3)} 
,
}
\]
where $\g{A}_h $, $\g{A}_c $ and $\g{A}_{ch}$ are deterministic real-valued, $M_2 \times M_2$, $N_1 \times N_1$ and $N_1 \times M_2$ dimensional matrices.

According to the assumptions of Proposition \ref{prop2} and the previous equations, the terms $(1)$, $(2)$ and $(3)$ can be rewritten:
\[
\Tab{rl}
{
(1)=&
\sum \limits_{i=1}^{M_2} \sum \limits_{j=1}^{M_2} \PP{\g{A}_h}_{ij} \PP{\g{h_2}\PP{\EntreeCodeDeux{}}}_i \PP{\g{h_2}\PP{\EntreeCodeDeux{}}}_j
\\[10 pt]
=&
\sum \limits_{i=1}^{M_2} \sum \limits_{j=1}^{M_2} \PP{\g{A}_h}_{ij} 
m_i\PP{\g{x}_2} m_j\PP{\g{x}_2} 
f\PP{\varphi_1,\PP{\g{\alpha}_i}_3,\PP{\g{\alpha}_i}_1,\PP{\g{\alpha}_i}_2} 
f\PP{\varphi_1,\PP{\g{\alpha}_j}_3,\PP{\g{\alpha}_j}_1,\PP{\g{\alpha}_j}_2},
\\[10 pt]
=&
\sum \limits_{i=1}^{M_2} \sum \limits_{j=1}^{M_2} \PP{\g{A}_h}_{ij} 
m_i\PP{\g{x}_2} m_j\PP{\g{x}_2} 
f\PP{\varphi_1,\PP{\g{\alpha}_i+\g{\alpha}_j}_3,\PP{\g{\alpha}_i+\g{\alpha}_j}_1,\PP{\g{\alpha}_i+\g{\alpha}_j}_2}
,
}
\]

\vspace{0.5cm}

\[
\Tab{rl}
{
(2)=&
\sum \limits_{i=1}^{N_1} \sum \limits_{j=1}^{N_1} \PP{\g{A}_c}_{ij} 
C_2\PP{\EntreeCodeDeux{},\EntreeCodeDeux{(i)}}
C_2\PP{\EntreeCodeDeux{},\EntreeCodeDeux{(j)}}
\\[10 pt]
=&
\sum \limits_{i=1}^{N_1} \sum \limits_{j=1}^{N_1} \PP{\g{A}_c}_{ij} 
\sigma_2^4 l\PP{\g{x}_2-\g{x}^{(i)}_2}   l\PP{\g{x}_2-\g{x}^{(j)}_2}
\\[10 pt]
& \qquad
\sum \limits_{n=1}^{n_{\varphi_1}} a_n f\PP{\varphi_1-\varphi^{(i)}_1,0 , \dfrac{-1}{2l^2_{1}},2n}
\sum \limits_{m=1}^{n_{\varphi_1}} a_m f\PP{\varphi_1-\varphi^{(j)}_1,0 , \dfrac{-1}{2l^2_{1}},2m}
\\[10 pt]
=&
\sum \limits_{i=1}^{N_1} \sum \limits_{j=1}^{N_1} \PP{\g{A}_c}_{ij} 
\sigma_2^4 l\PP{\g{x}_2-\g{x}^{(i)}_2}   l\PP{\g{x}_2-\g{x}^{(j)}_2}
\\[10 pt]
& \qquad
\sum \limits_{n=1}^{n_{\varphi_1}} \sum \limits_{m=1}^{n_{\varphi_1}}
a_n a_m  f\PP{\varphi_1-\varphi^{(i)}_1,0 , \dfrac{-1}{2l^2_{1}},2n}
 f\PP{\varphi_1-\varphi^{(j)}_1,0 , \dfrac{-1}{2l^2_{1}},2m }
,
}
\]

\vspace{0.5cm}

\[
\Tab{rl}
{
(3)=&
\sum \limits_{i=1}^{N_1} \sum \limits_{j=1}^{M_2} \PP{\g{A}_{ch}}_{ij} 
C_2\PP{\EntreeCodeDeux{},\EntreeCodeDeux{(i)}}
\PP{\g{h_2}\PP{\EntreeCodeDeux{}}}_j
\\[10 pt]
=&
\sum \limits_{i=1}^{N_1} \sum \limits_{j=1}^{M_2} \PP{\g{A}_{ch}}_{ij} 
\sigma_2^2 l\PP{\g{x}_2-\g{x}^{(i)}_2}  m_j\PP{\g{x}_2}

\\[10 pt]
& \qquad
\sum \limits_{n=1}^{n_{\varphi_1}} a_n f\PP{\varphi_1-\varphi^{(i)}_1,0 , \dfrac{-1}{2l^2_{1}},2n}
f\PP{\varphi_1,\PP{\g{\alpha}_j}_3,\PP{\g{\alpha}_j}_1,\PP{\g{\alpha}_j}_2}
.
}
\]

\vspace{0.5cm}

According to the fact that $m_i$ and $l$ are deterministic functions, $\g{x}_2$ and $\g{x}_2^{(i)}$ deterministic vectors, $\g{A}_h$, $\g{A}_c$ and $\g{A}_{ch}$ deterministic matrices, and $\varphi^{(i)}_1$ and $a_i$ deterministic real numbers, it can be written:
\[
\Mean{(1)}=
\sum \limits_{i=1}^{M_2} \sum \limits_{j=1}^{M_2} \PP{\g{A}_h}_{ij} 
m_i\PP{\g{x}_2} m_j\PP{\g{x}_2} 
\Mean{f\PP{\varphi_1,\PP{\g{\alpha}_i+\g{\alpha}_j}_3,\PP{\g{\alpha}_i+\g{\alpha}_j}_1,\PP{\g{\alpha}_i+\g{\alpha}_j}_2}}
,
\]

\vspace{0.2cm}

\[
\Tab{rl}
{
\Mean{(2)}=&
\sum \limits_{i=1}^{N_1} \sum \limits_{j=1}^{N_1} \PP{\g{A}_c}_{ij} 
\sigma_2^4 l\PP{\g{x}_2-\g{x}^{(i)}_2}   l\PP{\g{x}_2-\g{x}^{(j)}_2}
\\[15 pt]
& \qquad
\sum \limits_{n=1}^{n_{\varphi_1}} \sum \limits_{m=1}^{n_{\varphi_1}}
a_n a_m  \Mean{f\PP{\varphi_1-\varphi^{(i)}_1,0 , \dfrac{-1}{2l^2_{1}},2n}
 f\PP{\varphi_1-\varphi^{(j)}_1,0 , \dfrac{-1}{2l^2_{1}},2m}},
}
\]

\vspace{0.2cm}

\[
\Tab{rl}
{
\Mean{(3)}=&
\sum \limits_{i=1}^{N_1} \sum \limits_{j=1}^{M_2} \PP{\g{A}_{ch}}_{ij} 
\sigma_2^2 l\PP{\g{x}_2-\g{x}^{(i)}_2}  m_j\PP{\g{x}_2}

\\[10 pt]
& \qquad
\sum \limits_{n=1}^{n_{\varphi_1}} a_n 
\Mean{
f\PP{\varphi_1-\varphi^{(i)}_1,0 , \dfrac{-1}{2l^2_{1}},2n}
f\PP{\varphi_1,\PP{\g{\alpha}_j}_3,\PP{\g{\alpha}_j}_1,\PP{\g{\alpha}_j}_2}
}
.
}
\]

The means $\Mean{(1)}$, $\Mean{(2)}$ and $\Mean{(3)}$ can therefore be calculated analytically, and consequently, the mean\\
 $\Mean{\PP{\mu_2^{c}\PP{\EntreeCodeDeux{}}}^2+\PP{\sigma_2^{c}\PP{\EntreeCodeDeux{}}}^2 }$ can be calculated analytically.

\vspace{1cm}

From the two previous paragraphs and Proposition 1, it can be inferred that if verifying the assumptions of Proposition \ref{prop2}, then the first and the second moments of $\widehat{y}_{\text{nest}}^c(\g{x}_1,\g{x}_2)$ can be calculated analytically.

\vspace{1cm}

\subsection*{Proof of Proposition \ref{prop3}}

If $\widehat{y}_{\text{nest}}^c(\g{x}_1,\g{x}_2)=\widehat{y}^c_2(\widehat{y}^c_1(\g{x}_1),\g{x}_2)$ where $\widehat{y}_i^c=\mu^c_i+\varepsilon_i^c$, $\varepsilon_i^c \sim \text{GP}\PP{0,C_i^c},\ i\in\Ac{1,2}$, then if $\varepsilon_1^c$ is small enough, the process $\widehat{y}_{\text{nest}}^c(\g{x}_1,\g{x}_2)$ can be linearized:
\[
\begin{split}
\widehat{y}_{\text{nest}}^c(\g{x}_1,\g{x}_2) & =\mu^c_2(\mu^c_1(\g{x}_1)+\varepsilon^c_1(\g{x}_1),\g{x}_2)+\varepsilon^c_2(\mu^c_1(\g{x}_1)+\varepsilon^c_1(\g{x}_1),\g{x}_2), \\
& \approx \mu^c_2(\mu^c_1(\g{x}_1),\g{x}_2)
+\dfrac{\partial \mu_2^c}{\partial \varphi_1}(\mu_1^c(\g{x}_1),\g{x}_2)\varepsilon^c_1(\g{x}_1)
+\varepsilon^c_2(\mu^c_1(\g{x}_1),\g{x}_2),
\end{split}
\]

$\varepsilon_1$ and $\varepsilon_2$ being Gaussian processes, the predictor of the nested code can therefore be written as a Gaussian process:
\[
\widehat{y}_{\text{nest}}^c(\g{x}_1,\g{x}_2) 
\approx \mu^c_2(\mu^c_1(\g{x}_1),\g{x}_2)+\varepsilon^c_{\text{nest}}(\mu^c_1(\g{x}_1),\g{x}_2),
\]

\noindent{}where $\varepsilon^c_{\text{nest}}$ is a centred Gaussian process, whose covariance function, $C^c_{\text{nest}}$, is given by:
\[
\begin{split}
C^c_{\text{nest}}((\g{x}_1,\g{x}_2),(\g{x}_1',\g{x}_2')) 
& =C_2^c((\mu_1^c(\g{x}_1),\g{x}_2),(\mu_1^c(\g{x}_1'),\g{x}_2')) \\ 
+ & \dfrac{\partial \mu_2^c}{\partial \varphi_1}\PP{(\mu_1^c(\g{x}_1),\g{x}_2)}
\dfrac{\partial \mu_2^c}{\partial \varphi_1}\PP{(\mu_1^c(\g{x}_1'),\g{x}_2')}
C_1^c(\g{x}_1,\g{x}_1').
\end{split}
\]

\newpage

\bibliographystyle{plain}
\bibliography{biblio}

\begin{thebibliography}{10}

\bibitem{Bachoc2013}
F~Bachoc.
\newblock {\em {Parametric estimation of covariance function in
  Gaussian-process based Kriging models. Application to uncertainty
  quantification for computer experiments}}.
\newblock PhD thesis, {Universit{\'e} Paris-Diderot - Paris VII}, 2013.

\bibitem{Baker1977}
C.~T.~H. Baker.
\newblock {\em The numerical treatment of integral equations}.
\newblock Clarendon Press, Oxford, 1977.

\bibitem{Bect2012}
J.~Bect, D.~Ginsbourger, L.~Li, V.~Picheny, and E.~Vazquez.
\newblock {Sequential design of computer experiments for the estimation of a
  probability of failure}.
\newblock {\em Statistics and Computing}, 22:773--793, 2012.

\bibitem{Berger2001}
J.~O. Berger, V.~De~Oliveira, and B.~Sans\'o.
\newblock Objective bayesian analysis of spatially corellated data.
\newblock {\em Journal of the American Statistical Association},
  96(456):1361--1374, 2001.

\bibitem{Bichon2008}
B.~J. Bichon, M.~S. Eldred, L.~P. Swiler, S.~Mahadevan, and J.~M. Mcfarland.
\newblock {Efficient Global Reliability Analysis for Nonlinear Implicit
  Performance Functions}.
\newblock {\em AIAA Journal}, 46:2459--2468, 2008.

\bibitem{Chevalier2014}
C.~Chevalier, J.~Bect, D.~Ginsbourger, and E.~Vazquez.
\newblock {Fast parallel kriging-based stepwise uncertainty reduction with
  application to the identification of an excursion set}.
\newblock {\em Technometrics}, 56(4):455--465, 2014.

\bibitem{Echard2011}
B.~Echard, N.~Gayton, and M.~Lemaire.
\newblock {AK-MCS: An active learning reliability method combining Kriging and
  Monte Carlo Simulation}.
\newblock {\em Structural Safety}, 33:145--154, 2011.

\bibitem{Fang2006}
K.T. Fang, R.~Li, and A.~Sudjianto.
\newblock {\em Design and modeling for computer experiments}.
\newblock Chapman $\&$ Hall, Computer Science and Data Analysis Series, London,
  2006.

\bibitem{Fang2003}
K.T. Fang and D.K. Lin.
\newblock Uniform experimental designs and their applications in industry.
\newblock {\em Handbook of Statistics}, 22:131--178, 2003.

\bibitem{Ginsbourger2010}
D.~Ginsbourger, R.~Le~Riche, and L.~Carraro.
\newblock {\em Computational Intelligence in Expensive Optimization Problems},
  volume~2 of {\em Adaptation Learning and Optimization}, chapter Kriging Is
  Well-Suited to Parallelize Optimization, pages 131--162.
\newblock Springer Berlin Heidelberg, 2010.

\bibitem{Gramacy2012Tech}
R.~Gramacy and H.~Lian.
\newblock Gaussian process single-index models as emulators for computer
  experiments.
\newblock {\em Technometrics}, 54:1:30--41, 2012.

\bibitem{Gramacy2012Stat}
R.~B. Gramacy and H.~K.~H. Lee.
\newblock Cases for the nugget in modeling computer experiments.
\newblock {\em Statistics and Computing}, 22:713--722, 2012.

\bibitem{Hu2017}
R.~Hu and M.~Ludkovski.
\newblock Sequential design for ranking response surfaces.
\newblock {\em SIAM/ASA Journal on Uncertainty Quantification}, 5:212--239,
  2017.

\bibitem{Kennedy2000}
M.~C. Kennedy and A.~O'Hagan.
\newblock {Predicting the output from a complex computer code when fast
  approximations are avalaible}.
\newblock {\em Biometrika}, 87:1--13, 2000.

\bibitem{Kennedy2001}
M.~C. Kennedy and A.~O'Hagan.
\newblock {Bayesian Calibration of Computer Models}.
\newblock {\em Journal of the Royal Statistical Society. Series B (Statistical
  Methodology}, 63(3):425--464, 2001.

\bibitem{Kleijnen2017}
Jack~P.C. Kleijnen.
\newblock Regression and kriging metamodels with their experimental designs in
  simulation: A review.
\newblock {\em European Journal of Operational Research}, 256:1--16, 2017.

\bibitem{Stein1999}
Stein M.L.
\newblock {\em Interpolation of Spatial Data: Some Theory for Kriging.}
\newblock Springer, New York, 1999.

\bibitem{Paulo2005}
R.~Paulo.
\newblock Default priors for gaussian processes.
\newblock {\em Annals of Statistics}, 33(2):556--582, 2005.

\bibitem{Perrin2016}
G.~Perrin.
\newblock {Active learning surrogate models for the conception of systems with
  multiple failure modes}.
\newblock {\em {Reliability Engineering and System Safety}}, 149:130--136,
  2016.

\bibitem{PerrinSFDS2017}
G.~Perrin and C.~Cannamela.
\newblock A repulsion-based method for the definition and the enrichment of
  opotimized space filling designs in constrained input spaces.
\newblock {\em Journal de la Soci\'et\'e Fran\c{c}aise de Statistique},
  158(1):37--67, 2017.

\bibitem{PerrinJCP2017}
G.~Perrin, C.~Soize, S.~Marque-Pucheu, and J.~Garnier.
\newblock Nested polynomial trends for the improvement of gaussian
  process-based predictors.
\newblock {\em Journal of Computational Physics}, 346:389--402, 2017.

\bibitem{Rasmussen2006}
C.~E. Rasmussen and C.~K.I. Williams.
\newblock {\em Gaussian Processes for Machine Learning}.
\newblock The MIT Press, Cambridge, 2006.

\bibitem{Robert2007}
C.~Robert.
\newblock {\em The Bayesian Choice}.
\newblock Springer-Verlag New York, New York, 2007.

\bibitem{Sacks1989}
J.~Sacks, W.~Welch, T.~J. Mitchell, and H.~P. Wynn.
\newblock {Design and Analysis of Computer Experiments}.
\newblock {\em Statistical Science}, 4:409--435, 1989.

\bibitem{Santner2003}
T~J. Santner, B~J. Williams, and W~Notz.
\newblock {\em The design and analysis of computer experiments}.
\newblock Springer series in statistics. Springer, New York, 2003.

\end{thebibliography}

\end{document}